\documentclass[12pt,preprint]{aastex}
\slugcomment{Accepted for publication 
in {\it the Astrophysical Journal}  and it is scheduled for ApJ February 20, 2007, v 656, 2 issue} 
\def\lax {\ifmmode{_<\atop^{\sim}}\else{${_<\atop^{\sim}}$}\fi} 
\def\gax {\ifmmode{_>\atop^{\sim}}\else{${_>\atop^{\sim}}$}\fi} 

\begin{document}

\title{Effects of Downscattering   on the Continuum and Line Spectra in Powerful Wind Environment.  Monte Carlo Simulations, Analytical Results and Data Analysis}

\author{Philippe Laurent\altaffilmark{1,2} and Lev Titarchuk\altaffilmark{3,4}}

\altaffiltext{1}{CEA/DSM/DAPNIA/SAp, CEA Saclay, 91191 Gif sur Yvette, France; 
F\'ed\'eration APC, Coll\`ege de France, 75231 Paris, France;plaurent@cea.fr}
\altaffiltext{2}{F\'ed\'eration APC, Coll\`ege de France, 75231 Paris, France}
\altaffiltext{3}{George Mason University/Center for Earth
Observing and Space Research, Fairfax, VA 22030; and US Naval Research
Laboratory, Code 7655, Washington, DC 20375-5352; ltitarchuk@ssd5.nrl.navy.mil }
\altaffiltext{4}{NASA/ Goddard Space Flight Center, code 660, Greenbelt 
MD 20771, lev@lheapop.gsfc.nasa.gov}

\begin{abstract}
In Paper by Titarchuk \& Shrader  the general formulation and results for photon reprocessing (downscattering) that included recoil and Comptonization effects 
due to divergence of the flow were presented.  Here we show the Monte Carlo (MC) simulated continuum and line spectra. We also provide an analytical description of the simulated continuum spectra  using  the diffusion  approximation.
We have simulated the propagation of monochromatic and continuum  photons 
in a  bulk  outflow from a compact object. Electron scattering of the photons
within the expanding flow leads to a decrease of their energy which is of 
first order in $V/c$ (where $V$ is the outflow velocity).  The  downscattering effect 
of  first order in $V/c$  in the diverging flow is explained by semi-analytical calculations and confirmed by MC simulations. We  conclude that  redshifted lines and downscattering bumps  are intrinsic properties of the powerful  outflows for which  Thomson optical depth is greater than one. We fitted our model line profiles to the observations  using  four free parameters, $\beta=V/c$, optical depth of the wind $\tau$, the wind temperature $kT_e$ and 
the original line photon energy $E_0$. We show
how  the primary spectrum emitted close to the
black hole is  modified by reprocessing in the warm wind. In the framework of the our wind model  the fluorescent iron line K$_{\alpha}$  is formed in the partly ionized wind  as a result of illumination by central source continuum photons.  The demonstrated application of our outflow model  to  the XMM observations of MCG 6-30-15, 
and to the ASCA observations of GRO J1655-40,    points out a potential 
powerful spectral  diagnostic for probes of the outflow-central object connection in Galactic and 
extragalactic BH sources. 

 \end{abstract}

\section{Introduction}
Titarchuk \& Shrader (2005), Paper I, hereafter TSh05,  have shown  that downscattering modification of the primary photon spectrum by an outflowing plasma is a possible mechanism  for producing  the continuum excess in the $\sim10$ keV spectral region. This continuum excess can be formed only  if the energy index of the primary photon spectrum is less than one (see also Sunyaev \& Titarchuk 1980, hereafter ST80).  
This is usually attributed to Comptonization by a static reflector, such as a downward- or
obliquely-illuminated accretion disk,  although the overall continuum
form differs from that of the basic Compton reflection form. TSh05 thus suggest, that  in at least some cases, the outflow downscattering effect rather than the standard Compton reflection  mechanism is
responsible for the observed ''excess`` hard-X-ray continuum.   



To obtain  a description of the X-ray photon spectrum,  many  authors (see e.g. Gilfanov, Churazov \& Revnivtsev 1999, hereafter
GCR99, Pottschmidt et al. 2003, and etc) 
 used an empirical model in which each source spectrum is a sum of power law spectrum with photon index
$\Gamma$ and a multi-temperature disk blackbody. To this continuum, a reflection spectrum after Magdziarz  \& Zdziarski  (1995) was 
added. 
However,  Shaposhnikov \& Titarchuk (2006), hereafter ST06  argue that these spectral features are results of reprocessing  of the central
 source hard radiation in the outflow (warm wind) 
that surrounds the central BH. This  claim is supported by observed  power decay of quasiperiodic oscillations (QPOs) towards the softer states that is presumably a result of reprocessing of time signal in  the extended relatively warm wind (see Fig. 6 in ST06).   

The basic idea is that
electron  scattering of photons from a central source
entering the expanding outflow experience a decrease in
energy (downscattering). The magnitude of this decrease is of first order 
in $V/c$ and in $E/m_ec^2$ where $V$ is the outflow speed, 
c is the speed of light, $E$ is the initial photon energy
and $m_e$ is the electron rest mass.

TSh05 have developed an analytic formulation for the
emergent spectrum resulting from photon diffusion in
a  spherically expanding Comptonizing media,
characterized by two model  parameters: an average number of scatterings in the medium $N_{av}$, and 
efficiency the energy loss in the divergent flow $\varepsilon$.
In this formulation of the Radiative Transfer problem, the number of photons 
emitted in the wind shell equals that which escape to the Earth
observer  (i.e. the photon number is conserved). Consequently,  if   the high energy photons lose 
their energy on the way out but the number of photons is conserved, it  has to result in the accumulation of the photons 
at a particular lower energy band. 
 
 The basic idea of the wind photon  downscattering  is depicted in Figure 1. There  presents a simple explanation 
of the diverging flow  effect 
on the photon propagation through the medium. 
A photon emitted outwards near inner boundary and then 
scattered at a certain point by an electron moving with velocity ${\bf V}_1$, is
received by an electron moving with velocity ${\bf V}_2$ as shown with
frequency $\nu _2 = \nu _1\left[1+\left({\bf V}_1-{\bf V}_2\right)
\cdot{\bf n}/c\right]$ where ${\bf n}$ is a unit vector along the
path of the photon at the scattering point.  In a diverging flow
$\left({\bf V}_1-{\bf V}_2\right)\cdot{\bf n}/c <0$ and photons are
successively redshifted, until scattered to an observer at infinity.
 The color of photon path  (in Figure 1) indicates
the frequency shift in the rest frame of the receiver (electron or the Earth
observer). On the other hand, referring to the right-hand side of Figure 1, 
in a converging flow
$\left({\bf V}_1-{\bf V}_2\right)\cdot{\bf n}/c >0$ and photons are
blueshifted.

In this Paper we present the results of extensive Monte Carlo simulations of the downscattering effects in the
wind (diverging flow). In addition to the scattering, Doppler and Compton effects we include a photoabsorption effect in the diverging flow (wind).
Any photon with energy of about 7-8 keV and higher interacting with outflow plasma is more likely to be absorbed by the flow and be reemitted at energies about
6.4-6.6 keV, depending on the ionization stage of the flow (see Kallman et al. 2004). Laming and Titarchuk (2004), hereafter LT04, calculated the temperature and ionization balance
in the outflow determined by seeking the temperature at which the outflowing gas attains photoionization-recombination equilibrium.
The gas is heated by Compton scattering and photoionizations by photons from the central compact object and is cooled by radiation, ionization, 
and adiabatic
expansion losses. The solution of this  problem is  general regarding the calculation of the  temperature and ionization balance in the target
illuminated by the hard radiation. It potentially  has many applications beyond  Astrophysics
(for example, in   Radiation Safety Physics). Our simulations allow us to reveal {\it the spectral, timing and absorption properties of the emergent radiation}.

We  update the analytical results of the downscattering modifications of the continuum obtained by TSh05 (using Fokker-Planck technique) 
and we verify them using  the Monte Carlo (MC)
simulated spectra. We demonstrate that the  analytical formulas for the continuum spectra well describe
 the MC spectra almost for any $\beta=V/c\lax 0.8$. 
  
   
 
Many authors [e.g. Tanaka et al. (1995); Nandra et al. (1997); Wilms et al. (2001); 
Uttley et al. (2004); Miller et al. (2004a,b)]
have found unusual curvature (red-skewed  features) at energies between 2 and 8 keV in X-ray spectra of a number of 
galactic and extragalactic BH sources. Significantly the fluxes of these features remain nearly {\it constant} 
despite the large changes in
the continuum flux as shown in the observations of MCG-6-30-15, NGC 4051 and others [Markovitz,
Edelson, \& Vaughan (2003)]. This lack of variability means that fast changes (which are almost certain to occur near
the BH) are not obtained.
These observations motivate us to make a detailed study of the spectral line formation in the outflows that  possibly occur in black hole sources.
In this Paper we  demonstrate that  red-skewed  features can be a  result of scattering of iron line photons  in the powerful outflow.

We offer a model in which the hard radiation of the central object illuminates and heats the outflow (wind) region
originated in the outskirts of the disk (well outside the innermost part of the accretion disk near the BH).  
The line photons are generated and scattered in the
outflow. 
Our basic scenario is illustrated in Figure 1. The
wind originates at a distance $r_{in}$ from the central black hole
and is of a density such as to give a Thomson scattering optical depth $\tau_{\rm T}$
close to unity far from the black hole. The optical depth in the Fe K
continuum is about 1-3 times higher than that due to electron scattering
(assuming a solar abundance of Fe, and depending on charge state, see
Kallman et al. 2004),
and so Fe K$\alpha$ formed by inner shell ionization of Fe ions in the
outflow by the continuum from the central black hole only comes from a
smaller inner region.
In fact,  using the calculations of the  ionization state of the outflow shell 
by  LT04,  we find that K$_\alpha-$photon sources are exponentially
distributed  over  a skin at the bottom of the outflow, the skin thickness being $4\%$ of the optical depth of the wind shell. 
Furthermore because
the iron photo-absorption opacity $\sigma_K$ is about few times
the Thomson opacity, $\sigma_{\rm T}$, any photon 
around the K-absorption edge energy must be absorbed. In fact, these edges are seen with the strong
K$_\alpha$ lines during  X-ray superbursts detected from NS source 4U 1820-30 (see  Figs. 5 and 9 in Strohmayer \& Brown 2002).

A thorough analysis  and review of the diffusion theory of photon propagation 
in an optically  thick fluid in bulk outflow  
has been provided by Titarchuk, Kazanas \& Becker (2003), hereafter TKB.  
They show that  the iron line is produced in an effectively optically thick medium.
Its red wing is the result of multiple scattering,  each scattering producing a first
order $V/c-$redshift. This process produces a red wing
to the line without any particular fine-tuned geometric arrangement.

  The TSh05 and TBK results are obtained using the Fokker-Planck equation [Blandford \& Payne (1981) see
also Lifshitz \& Pitaevsky (1981) for a general treatment of the particle acceleration problem]
which   is valid for $\beta=V/c\ll 1$ and $\tau_{eff}>1$. The MC
method does not have these limitations.  

We show the results of MC simulations of scatterings of continuum and monochromatic photons  
for various values of plasma temperatures $kT_e$, outflow bulk velocity  $\beta$, and Thomson optical depth 
 $\tau$. The details of the MC method applied to the outflow model are given  
in \S 2.    We  compare the results of the MC simulated continuum with their analytical description and discuss the downscattering modifications 
of the incident spectrum in \S \S3.1-3.2  The results of  the simulated line formation in the outflow and its explanation 
using the analytical theory are shown in \S 3.3.
 Examples of the model fits to the XMM data obtained for MCG-6-30-15 
and to ASCA archive data of galactic BHs
are shown in \S  4. 
The results of the simulations of overall spectra including the scattering, photon absorption effects are presented in \S 5.
Discussion and conclusions follow in \S 6.


\section{Details of the Monte Carlo Method}
Laurent \& Titarchuk (1999), hereafter LT99, simulated the Comptonization of photons 
using a Monte-Carlo method in the  Schwarzschild metric. 
The geometry used in these simulations consists of a spherical shell harboring a black hole 
in its center.  Once given the cloud Thomson optical depth $\tau$,  
 $N_e$ is an electron number density measured in 
the local rest frame of the flow that is  
\begin{equation}
N_e=\tau (r_{\rm S}/r)^{2}/[r_{\rm S}\sigma_T(r_{\rm S}/r_{in} - r_{\rm S}/r_{out})]. 
\label{electrden}
\end{equation}
Here $\sigma_T $ is the Thomson cross section, 
$r_{in}$ and $r_{out}$  are the inner and outer radii of the shell respectively, 
and $r_{\rm S}$ is the Schwarzschild radius. 
Simulations 
made for different values of  $r_{in}$ and $r_{out}$ (that are much greater than the Schwarzschild radius) have shown 
no change of the line profile and continuum
provided the shell optical depth remains the same.
We have varied  the shell electron temperature from 0.01 to 1 keV. 
Along with the thermal motion of the electrons in the cloud,  
we have also taken into account the bulk velocity of the outflow $\beta=V/c$.
In these simulations, the spherical wind has a constant bulk velocity within the flow.
It is worth noting that an increase of the shell temperature from 0.01 to 1 keV 
 slightly increases the blue wing of  the observed line, but induces no change in the red wing  that  
we are mostly interested in here. 

For the timing studies described in the next section
we have also computed   the path length  that each photon  in the simulations  has 
covered until it escapes from the shell. 

\section{Dowscattering effect in an outflow}
The problem of photon propagation in a fluid in bulk motion has been studied in
detail in a number of papers [see e.g.  Blandford \& Payne (1981),
 Payne \& Blandford (1981),  Nobili, Turolla
\& Zampieri (1993); Titarchuk, Mastichiadis \& Kylafis (1997); TKB; TSh05; Laurent \& Titarchuk (1999), (2001).
In particular, TSh05 present a general formulation and a solution of the spectral
formation in the diverging outflow. They demonstrated
that the resulting spectrum can be formed as a convolution of energy and spatial diffusion
solutions. 


To apply these Radiative Transfer results to observational data,
TSh05 have developed a generic analytical formulation which leads to a simple
analytic expression for modification of the emergent 
spectrum due to recoil and velocity 
divergence effects in the flow.  For the recoil effect, TSh05 extend 
the  results by Sunyaev \& Titarchuk (1980), hereafter ST80, 
 who show that the downscattering feature, or "bump", can
appear superposed on power-law spectra with energy spectral indices $\alpha<1$  
as a result of photon
diffusion through a static cloud.  From ST80 and  calculations presented in TSh05,
one might conclude such bumps are not a necessarily a feature of
disk reflection, but they can  be also formed as a result of the photon 
reprocessing  in a relatively cool, ambient plasma characterized by 
temperatures of order $10^6$ K.

\subsection{Time  distributions of escape photons}
In order to understand and explain  the shapes and  the main features of the MC simulated spectra  
we attempted to reproduce them using the analytical diffusion approximation. 
In framework  of the diffusion theory (TSh05) the resulting spectrum can be presented as a  particular convolution of a primary photon spectrum  $\varphi(z)$ and a time  distribution of escape photons $ {\cal P}(u)$ (see Eq. \ref{emerg}, below).  Here 
$u=t/t_{fp}$,  $t$ and 
$t_{fp}=l/c$ are the time for photon to escape from the wind  and the mean free path time between two consequent photon scatterings respectively; $z=E/m_ec^2$  is the dimensionless photon energy.
ST80 and Sunyaev \& Titarchuk (1985) hereafter ST85, demonstrate that the photon escape distribution ${\cal P}(u)$  exponentially decays with $u$  as $u>1$  for any bounded configuration, namely ${\cal P}(u)\sim b\exp(-bu)$. 
 
In Figure 2 we present the simulated  time distribution ${\cal P}(t/t_{fp})$ for 
optical depth of the wind $\tau$  equal to 4.
As it was found by ST80 and ST85, the time distribution decay exponentially  $\exp (-bt/t_{fp})$ for  $t>t_{fp}$ when the photons propagate in any bounded medium. 
This result is used for derivation of the analytical spectra (see Eqs \ref{emerg}). 
It means that the best-fit $b-$ parameters obtained by fitting  $\exp (-bt/t_{fp})$ to the time distributions and 
those $b-$ parameters obtained by fitting the analytical spectra to the simulated spectra 
 are identical within errors  of $b$.
We computed  the $b-$values obtained by these ways and we found they are  identical. This consistency check gives us more confidence of accuracy of our simulations and their analytical description.

In Figure 3  we show the photon distribution for emergent photons over a range of deflecting angles with respect of the  direction of the incident photons.
The outflow shell is radially illuminated be the central source radiation. 
The simulations are presented for $\tau=1, ~2,~4$ and  $\beta=0.1$. 
In fact, we find in our simulations that  the distribution of the scattering component   weakly depends on  $\tau$.
 Practically, there is little difference
between that related to $\tau=4$, and $\tau=1,~2$. There is only the difference in  the normalization  of the direct (non-scattering) component $A_N$ for which  the ratio of $A_N(\tau=4)/A_N(\tau)=\exp(-4+\tau)$.
The distribution has a strong maximum about $50^{o}$. Thus  we expect that our results for 
spherical wind can be applicable for the conical wind if the half-angle of the conical wind is about
$50^o$ or more  (wide open jet or wind).   

As we demonstrate in Figure 2  the simulated  escape time distribution in the wind  has an exponential profile which is a typical signature of the photon diffusion propagation in any bounded configuration (ST85). This leads us to conclusion that  we can also treat the radiative transfer in the wind in the framework of the diffusion model (see TSh05).   

\subsection{Monte Carlo simulated spectra of the continuum and their analytical description}
For the completeness of the present work  we reproduce some necessary details of the analytical diffusion model  by TSh05.
 Particularly,  the downscattering modification of the incident spectrum $\varphi(E)$ can be obtained using formula (24) in TSh05:
%
\begin{equation}
{\cal F}_\nu(z,\varepsilon)=\frac{1}{z}
\int_0^{u_{max, \varepsilon}(z)}e^{-4\varepsilon u/3}
\psi^{(\varepsilon)} (z,u)
\varphi[\psi^{(\varepsilon)}(z,u)] {\cal P}(u)du, 
\label{emerg}
\end{equation}
where ${\cal P}(u)$ is  the distribution of the photons over their dimensionless escape time $u=t/t_{fp}$,  
\begin{equation}
u_{max, \varepsilon}(z)=(3/\varepsilon)\ln(1+\varepsilon/3z)\gg 1, 
\label{umax}
\end{equation}
is  a maximal number of scatterings related to the dimensionless energy $z=E/m_ec^2$.
\begin{equation}  
z_0=\psi^{(\varepsilon)}(z,u)=(\varepsilon/3)/[(1+\varepsilon/3z)\exp(-\varepsilon u/3)-1], 
\label{z0}
\end{equation}
 is the dimensionless photon energy $z_0=E_0/m_ec^2$ at $u=0$ 
 and $\varepsilon\ll1$ is a coefficient of the diverging term  in the kinetic (Fokker-Planck) equation 
(see Eqs. 16, 17 in TSh05 and related  discussion). In fact, $\varepsilon$ is an average relative energy redshift per scattering.
Below (see Eq. \ref{deltaE}), we evaluate these values and verify them using results of   our Monte Carlo simulations.  
We find that $\varepsilon\sim \beta/2\tau$ is of  order of $10^{-2}$  for $\beta\lax0.1$ and for  $\tau$ of a few. 


This result is a generalization of ST80 and TKB results. ST80 derived the 
spectra when the downscattering effects  due to the recoil was taken into account.
It should be noted that the downscattering modification of the spectrum occurs
when photons undergo multiple scatterings $u$.
Thus  we  can expand the integrand function ~~~ 
$W^{(\varepsilon)}(z,u)=e^{-4\varepsilon u/3}[\psi^{(\varepsilon)} (z,u)]\varphi[\psi^{(\varepsilon)}(z,u)]$ over $u$
in formula (\ref{emerg}): 
 \begin{equation}
W^{(\varepsilon)}(z,u)\approx W^{(\varepsilon)}(z,0)+W^{(\varepsilon)\prime}_{u}(z,0)u +
W^{(\varepsilon)\prime\prime}_{u}(z,0)u^2/2 +W^{(\varepsilon)\prime\prime\prime}_{u}(z,0)u^3/3! +...
 \label{expan}
\end{equation}

 
For the simplest case of the power-law incident spectrum and neglecting the diverging term  in the kinetic (Fokker-Planck) equation 
($\varepsilon=0$) the resulting spectrum (Eq. \ref{emerg}) is as follows
\begin{equation}
{\cal F}_E(z)= \frac{A}{z}\int_{0}^{1/z}(1/z-u)^{\alpha-1}{\cal P}(u)du.
\label{powerlaw}
\end{equation}  
Using the expansion (\ref{expan}) one can obtain (see Appendix A, Eq. \ref{powerlaw_ap3})
$$
{\cal F}_E(z)= Az^{-\alpha}[ M_{0}(z^{-1})+
\frac{(1-\alpha)}{1!}z M_{1}(z^{-1})+\frac{(1-\alpha)(2-\alpha)}{2!}z^2 M_{2}(z^{-1})+...
$$
\begin{equation}
+\frac{(1-\alpha)(2-\alpha)...(n-\alpha)}{n!} z^{n}M_{n}(z^{-1}) +...],
\label{plexpan}
\end{equation}
 where 
\begin{equation}
M_{n}(z^{-1})=b\int_0^{1/z}u^n\exp(-bu)du ~~~~~{\rm~for}~~n=0,~1,~2~... 
\label{M_func}
 \end{equation}
Note, the first term in the parenthesis of the right hand side $M_{0}(z^{-1})=1$ for $z\ll 1$ and the second term of that   $(1-\alpha)z M_{1}(z^{-1})$ is positive if energy index of the primary photon spectrum $\alpha<1$. Thus one can expect the photon accumulation bump at low energies for the downscattered power-law spectra for which the  primary photon spectrum  is quite hard (the energy index  $\alpha<1$). Indeed, in the pure scattering  case the photon number is conserved,  so if large   numbers of   the downscattered high energy photons  are removed from the high energy part of the incident spectrum, then they should are seen at lower energies as a so called {\it  downscattering} bump.

In Figures 4, 5 we present the Monte Carlo simulated spectra along with the analytical spectra  (Eq. \ref{plexpan}) for the power-law incident spectra  for $\tau=4$ and  $\tau=2$ respectively. 

The downscattering (accumulation) bump and softening of spectrum at high energies are clearly seen in these spectra. As  optical depth increases   more  prominent  bumps are formed in the outflow. 
The softening of the analytic  spectra due to the downscattering effect is well described by exponential decay of  moments of the time distribution, 
$M_i(1/z)$ $(i=0, 1, 2,..)$ at higher energies (i.e. when $b/z>1$),  see Eqs. (\ref{M_func}), (\ref{M_n}).
The parameter $b$  is only a free parameter to fit the analytical formula,  we fix $\varepsilon=0$
for which $u_{max}=1/z$, see Eq. (\ref{umax}), namely $\lim_{\varepsilon\to0}u_{max}^{\varepsilon}=1/z$. 
For $\tau=4$ the analytical spectra (\ref{plexpan}) perfectly represent the simulated spectra for a wide range of the wind 
velocities $V=\beta c\lax0.8c$.  It is worth noting that we need not more than 6 terms   of asymptotic series (\ref{plexpan}) to achieve the highest accuracy   of the analytical spectra calculations.

For $\beta = 0.05,~0.1,~0.3,~0.8$ the best-fit parameters $b=0.07,~0.07,~0.11,~0.45$ respectively. The inverse of $b-$parameter, $N_{av}=1/b$ is an average
number of scatterings undergone by the photons in the wind.  For a faster wind the number of scatterings  naturally decreases because  photons are 
effectively taken by the outflow and less photons are scattered off outflow electrons.  ST85, calculate b-parameters (and the 
average number of scatterings $N_{av}$) for spherical and plane (disk) geometry (Table 1, in ST85, where $\beta$ corresponds to our value $b$). 
ST85 present their results for the static case i.e the zero bulk velocity case. Thus we should compare  their results with our results when
$\beta=0.05,~0.1$. Moreover, our wind extended envelope of optical depth  $\tau$ is equivalent to the disk of  half-optical depth  $\tau/2$.
For example, for $\tau=4$, our $b=0.07$ should be related to $0.101$ in ST85 for $\tau=2$. As expected our b-value is smaller (or $1/b$ greater)
than that in ST85, because the empty internal cavity of the wind envelope essentially  amplifies the number of photon  scatterings with respect 
to that in a plane disk of the same optical depth.   The scattering effect of the cavity is suppressed when the wind velocity (or $\beta$) increases.
For $\beta=0.3$, the cavity effect and the wind velocity on scattering compensate each other. Our value of $b=0.11$ is very close to that in ST85.


In Figure 5 we show the simulated spectra along with the analytical spectra for $\tau=2$. One can notice that
the best-fit analytical spectra  slightly deviate from  the simulated spectra at high energies. 
For $\beta = 0.05,~0.1,~0.3,~0.8$ the best-fit parameters $b=0.13,~0.15,~0.21,~0.62$ respectively. ST85 give the corresponding value of $b=0.234$
(related to our $b$-parameter for $\beta=0.05$). 
From this comparison it is evident that the scattering effect of the cavity becomes stronger with the optical depth decrease. 
As expected, the analytical spectra as solutions of the diffusion problem in Thomson regime may depart
 from the Monte Carlo simulated spectra for $\tau\lax2$ at high energies where the electron opacity is less than the Thomson opacity.  
The central source spectrum (dashed line) is slightly less modified by the scattering in the wind (see MC histogram for the simulated spectrum) than that using the diffusion spectrum (solid line).

Another  interesting case is the downscattering modification of Comptonization spectrum (see e.g. ST80 and Titarchuk
1994). The incident Comptonization  spectrum can be well fitted by a power-law spectrum with an exponential cutoff, namely
 \begin{equation}
\varphi_{comp}(z) \approx Az^{-\alpha}\exp(-z/z_{\ast}).
 \label{comp}
\end{equation}
where  $z_{\ast}=E_{\ast}/m_ec^2$ is the dimensionless photon energy and 
the cutoff energy $E_{\ast}$ is related to the Compton cloud electron temperature $kT_e$,
namely $E_{\ast}\approx 2kT_e$. 
The downscattering modification of the  spectrum, (see Eq. \ref{intexpan1}) can be written  using the expansion (\ref{expan}).  
It can be shown that for $\varepsilon=0$ and $z_{\ast}\to\infty$ formula (\ref{intexpan1}) is reduced to formula (\ref{plexpan}).
In Figures 6 and 7 we present the Monte Carlo simulated spectra for the power-law with exponential-cutoff incident spectra for $\tau=4, ~\tau=2$
respectively  along with the analytically calculated spectra 
(see Eqs. \ref{intexpan1}, \ref{nder}-\ref{highz_0vsdu}). 
 The  MC simulated spectra are  well described by the analytical  model for all $\beta\lax0.8$. It is worth noting that  the downscattering bumps are not seen in either the simulated nor model spectra.

For $\tau=4$ and  $\beta = 0.1,~0.3,~0.5, ~0.8$ the best-fit parameters are $b=0.06,~0.10,~0.18,~0.38$ respectively.
For $\tau=2$ and  the same  $\beta$ the best-fit parameters are $b=0.14,~0.21,~0.30,~0.59$ respectively.
It is worth noting that the values of $b-$parameter for small $\beta\lax0.1$ is almost independent of the shape of the incident spectrum 
(within the error bar
$\delta b=\pm 0.005$)  and it depends on $\tau$ only.

Comparison of the simulated and analytically calculated spectra leads us to conclude that the simulated spectra can be well described by the analytical model  with only one  free parameter $b$ that is related to the mean number of photon scattering in the wind ($N_{av}=1/b$).
It means that the continuum spectral formation in the pure scattering wind is dictated  by the mean  number of scatterings  $N_{av}$ only but that
the value of $N_{av}$ is determined by the combined effect of the wind optical depth $\tau$ and the dimensionless wind velocity $\beta$.       



\subsection{ K$_{\alpha}-$Line formation in the outflow}
   
Another observational feature of the wind is the strong broad feature of K$_{\alpha}$ line in
the spectrum.  Recently,  Shaposhnikov \& Titarchuk (2006) demonstrate  that these strong iron lines are present in all spectral states of Cyg X-1. Also they  show  that the equivalent
width (EW) of the  K$_{\alpha}$ line increases with the photon
index $\Gamma$ from about 150 eV in the low/hard state to about 1.3 keV in the high/soft and very soft states.

The X-ray photons of the central source 
illuminates the wind (see Fig. 1). The wind gas is heated by Compton scattering and photoionizations by photons coming from the central object and is cooled by radiation, ionization,
and adiabatic expansion losses (LT04). 
The photons above the K-edge energy are absorbed and  ionize iron atoms which leads to the formation of
the strong K$_{\alpha}$ line.  LT04 calculated ionization, temperature structure and the equivalent widths of Fe K$_{\alpha}$ line formed in the wind. 
For the wide set
of parameters of the wind (velocity, the Thomson optical depth $\tau$) and the incident Comptonization spectrum 
(the index and the Compton cloud electron temperature) they established 
that EW of the line should be about  1 keV and less for the line to  be observed. 
LT04 also predicted that in this case the inner radius of the wind should be situated at $(10^{3}-10^{4})/\tau_{0}$ Schwarzschild radii away from the
central object.

\subsubsection{The spectral line redshift in outflow. Analytical Description}
  
In Figure 1 we show the picture of the line photon propagation in the wind.
At each consecutive scattering the line photon in average loses its energy, in other words it
is redshifted.  
A photon emitted near the inner boundary and subsequently scattered
 by an electron moving with velocity ${\bf V}_1$, is
received by an electron moving with velocity ${\bf V}_2$ as shown in Fig. 1. 
The change in frequency is
\begin{equation}
 \nu _2 = \nu _1\left[1+\left({\bf V}_1-{\bf V}_2\right)
\cdot{\bf n}/c\right]
\label{energ_change}
\end{equation}
where ${\bf n}$ is a unit vector along the
path of the photon scattered at the next point. 

We remind a reader that   in the static and spherical-symmetric  medium  and 
where electrons experience a random (Brownian) motion only,   the  mean energy change per scattering, $<\Delta E>$, for photons undergoing quite  a few scatterings is proportional to   square  of electron velocity, namely 
$<\Delta E>\propto (V/c)^2$  (see e.g. Rybicki \& Lightman 1979 and ST80). 
 
On the other hand   for photons undergoing numerous 
 scatterings in the bulk flow $<\Delta E>$ is just  proportional  to $V/c$. 
[see Titarchuk, Mastihiadis \& Kylafis 1997, Appendix D and TKB). 
In order to make estimate of the  loss per scattering, one has to find the response of the energy operator of the diffusion kinetic equation (see TKB, Eqs. 3, 18) to a delta-function injection. In Appendix B1 we show that in this  standard diffusion approximation $<\Delta E>$ is proportional to the velocity divergence, namely   
\begin{equation}
<\Delta E>\approx -\frac{4}{3}\frac{\nabla ({\bf V}/c)}{\kappa}E_0\sim \frac{8}{3}\frac{\beta}{\tau}E_0 
\label{deltaE_divergence}
\end{equation}
where $\kappa$ is the scattering coefficient (or the inverse of the scattering mean free path).
But this method  provides rather a qualitative estimate of the redshift  effect in the outflow than a quantitative one. 
The   value of the  numerical factor $4/3$  in equation (\ref{deltaE_divergence}) is not precisely correct,  which suggests that factor $4/3$ should be replaced by the right one; however, the form of the exact equation actually depends on $\beta=V/c$ and $\tau$ in a more complex way.
Formula (\ref{deltaE_divergence}) should be then taken only as an estimate.

A useful result of the  $ <\Delta E>$ estimate is obtained using the scattering geometry method  (see Appendix B2)
\begin{equation}
<\Delta E>\approx -\beta f/(2\tau)\sqrt{1-(f/2\tau)^2} E_0.
\label{deltaE}
\end{equation}
Below  we utilize  the MC simulations to show that   the numerical factor  $f$ in formula (\ref{deltaE}) is   about one.
Comparison of Eqs. (\ref{deltaE_divergence}) and (\ref{deltaE})  leads us to conclude that $<\Delta E>$ is proportional 
to $\beta$ but that is overestimated in equation   Eq. (\ref{deltaE_divergence}).   
 
The average energy of photons escaping after $N_{av}=1/b$ scatterings is  
\begin{equation}
<E>_{sc}=(1+<\Delta E>)^{1/b}E_0.  
\label{E_av_p}
\end{equation}
From here using formula (\ref{deltaE}) we obtain that
\begin{equation}
<E>_{sc}=\{1-\beta f/(2\tau)\sqrt{1-(f/2\tau)^2}\}^{1/b} E_0.
\label{E_av}
\end{equation} 

\subsubsection{MC simulated spectral lines in outflow. Evidence of the strong redshift effect in the outflow}
 
We simulate line profiles for different optical depths of the wind.  
In Figure 8 (left hand side panel) we present  examples of the line profiles for 
simulations where  we take into account only the pure electron scattering
in the flow, photo-electric absorption being neglected in these simulations. 
In Figure 8 (right-hand side panel) we present the spectral profile of the line obtained as a result of the electron scattering
in the flow along with   photo-electric absorption.   
When the X-ray  continuum spectrum of the central source illuminates the wind  the seed  $K_{\alpha}-$photons are formed in the innermost part   of the wind.
The photo-absorption profile as a function of energy, calculated by LT04, is related to the specific case of $\beta=0.1$,
  $\alpha=0.5$, $kT_{e}=50$ keV  for which
 the inferred LT04 model parameter $r_{in}\tau/L_{40}=6\times 10^{12}$ cm.
 The model and X-ray continuum parameters are the following: $r_{in}$ is the outflow inner radius, $L_{40}$ is the luminosity of the central source in units $10^{40}$ 
 erg s$^{-1}$, $\alpha$   is the energy spectral index and  $kT_{e}$ is 
 the electron temperature of the Comptonization spectrum 
 (see details of the photo-absorption calculations in LT04). We find that the height scale of 
 the source exponential distribution, $H$, is only  a small fraction of the inner radius or in other words 
   $H/r_{in}\sim0.04\tau$ is a small fraction of the  optical thickness of the outflow.    
Thus {\it the sources of K$_\alpha$ is distributed in the innermost part of the flow}. 
We simulate  the line profiles  for  $\tau=2, ~4$, 
$kT_e = 0.1$ keV,  $\beta =0.1$, and
for an initial seed photon energy $E_{ph} = 6.6$ keV.

One can clearly see  the red-skewed part of the spectrum formed by photons undergoing multiple scatterings, along with the 
primary peak formed by photons escaping directly to the observer. 
The normalization of the direct non-scattering line component is suppressed for $\tau=2$. The real normalizations are 
 $A_N=1.35\times10^6, ~1.29\times10^6$ for the diffusion and diffusion+absorption cases respectively.

In a  situation, when the effect of the photon photoionization and absorption is taken into 
account, the red wing of the line (for energies less than 4 keV) and   blue wing (for energies higher 8 keV)
have  to be suppressed  due to high photo-electric absorption 
opacities.  Despite  this photo-absorption suppression, the red-skewed shape of the line  is retained.
Thus one can conclude that {\it the red-skewed line profiles are intrinsic signatures of outward propagation of K$_{\alpha}-$photons 
generated at the bottom of the outflow due to the outflow illumination by the X-ray  hard radiation of the central source.}


In Figure 9  we show  timing properties of the line propagation in the  outflow 
for 5 energy bands for the pure scattering case.
In the plot the time is given in the free path  time units $t_{fp}=l/c$. 
Photons escaping from the bottom atmosphere without scattering are at 6.6 keV. 
The model parameters used here are  
$kT_e = 0.1$ keV, ~$ \beta = 0.1$, $\tau =4$ (upper panel),   and  $\tau=2$ (lower panel).
We show the arrival time of photons at the top of the cloud for five energy bands, 
in free path  time units $t_{fp}$.  Soft lags, low energy emission at later times are clearly seen. 

It  is also worth noting that TKB predict  soft lags, namely that a photon of initial energy $E_0$ loses its 
energy with time, i.e. $E=\exp(-u_{\rm TKB})E_0$ [compare with our formulas (\ref{deltaE}, \ref{E_av}) ]. The dimensionless time variable 
(we call it as $u_{\rm TKB}$) used there is  a mathematical variable for a convolution method (see their Eq. 9). It can be interpreted as 
a product of  the  number of scatterings $t/t_{fp}$ multiplied  by the mean photon inelasticity $<\Delta E>/E$ 
(see our Eq. \ref{deltaE}), i.e 
 $u_{\rm TKB}= (<\Delta E>/E)(t/t_{fp})$. 

The photon distribution  with respect to  escape time integrated over energies can be fitted  with an exponential 
law  $b\exp(-bt/t_{fp})$ (shown in Fig. 2). The values of $b=0.057$ and $b=0.144$ for $\tau=4$ and $\tau=2$ respectively 
(and consequently the average number of scatterings $N_{av}=1/b$) are very close to $b$ (or $N_{av}$) obtained 
from the best-fit analytical spectrum for the same values of $\beta$ and $\tau$. 

We also check  the analytical estimate of the average redshift of the line energy using formula (\ref{E_av}) and  
that obtained from our simulations with the  values  $<E>_{sc}=4.56$ keV and $<E>_{sc}=5.27$ keV  for $\tau=4$ and $\tau=2$ respectively.
On the other hand keeping in mind the values of $b$,  related to a given $\tau$ we can obtain the same values of $<E>_{sc}$ as the simulated ones using formula
(\ref{E_av}) with  $f=1.3, ~f=1.7$  for $\tau=2$ and $\tau=4$ respectively. 
These values  of $f>1$  are expected in the framework of our analytical  estimate of $<E>$.  
 (see details of the derivation in Eq. (\ref{l_r}).  
 It is worth noting that now we can  calculate $\varepsilon=<\Delta E>/E_0$
that follows from formula (\ref{deltaE}) using $f$ inferred from our simulations. 
They are $\varepsilon=0.03,$ and $0.04$ for $\tau=2,~4$  respectively and for $\beta=0.1$.    



In Figure 10 we present the emergent line profiles for   different values  of $\tau$ and $\beta$.
Spectral lines as a function of $\tau$ are for $\beta=0.1$, and that as a function of
$\beta$ for $\tau=2$.
The source photons are generated at the bottom of the cloud and the cloud temperature is kept fixed at $kT_e=0.1$ keV.
The red wing of the line profile is  insensitive to the temperature of the flow if $kT_e< 1$ 
keV. On the other hand, the blue wing becomes broader with $kT_e$,  but its relative width is still  smaller than that of 
the red wing. 
In fact, the blue wing  is affected for energies higher than K-edge ($>8$ keV)
if  the photo-absorption is taken into account (see  the left lower panels of Fig. 10). In general, 
the photo-absorption narrows the line profile 
(compare the lines profiles in the upper and lower panels of Fig. 10).   

The line profiles  depend quite strongly on $\tau$ and $\beta$.  
The shape of the red wing below  the broad peak (see  Fig. 10)
follows a power law with  index which is a strong function of $\tau$ and $\beta$. 
Our calculated indices are in agreement with 
the TKB results for $\beta<0.5$ and $\tau_{eff}=3\beta\tau>1$.   In TKB, the photon indices are equal to $\Gamma=2+\lambda_k^2$,
where $\lambda_k^2$ are eigenvalues  calculated using their equation (30). 

When $\beta\sim 1$  
the spectrum of scattered photons becomes flat even for $\tau=2$. 
The relativistic spherical outflow completely washes out the bump induced by scatterings (see the right panels of Fig. 10). 
However, for  relatively small $\beta$, the power law tails still remain steep for higher $\tau$ (see the left panels of Fig. 10).  
This is very similar to the results obtained by TKB.

\section{Comparison of MC simulated line profiles with the data}

In Figures 11 and 12 (the left panel) we present the results of  fitting our model to the XMM (EPIC) data of Wilms et al. (2001) obtained during the observation of 
the bright Seyfert 1 galaxy MCG-6-30-15. These observations were focused on the broad Fe K$\alpha$ line at
$\sim 6.4$ keV. For MCG 6-30-15   we fit the continuum with a power law of 
index 1.8 and compare the residuals with our model. 
The best-fit  parameters of the pure scattering model we found are $E_{ph} = 6.51$ keV, ~$kT_e = 0.1$ keV, 
$\tau =1.2, ~\beta = 0.02$. We generated the primary photons  at the bottom of outflow. If we take into account the LT04 model for  the photo-absorption along with
electron scattering, then the best-fit parameters become $E_{ph} = 6.51$ keV, ~$kT_e = 0.1$ keV, 
$\tau =1.7, ~\beta = 0.02$. 

 The best-fit model spectrum is very similar to the steep spectrum presented in Fig. 10 for $\beta=0.01$ (black curve in the
 lower panel). The quality of the fits  strongly depends on the outflow optical depth $\tau$ and bulk velocity
$\beta$. It is less sensitive to the outflow temperature $kT_e$ unless $kT_e\lax0.1$ keV.  
It is important to note that  using our diverging outflow model we confirm  Fabian's (1994) finding
that velocity of the warm absorber outflow is about 5000 km s$^{-1}$ in MCG-6-30-15. 


We fitted the red-skewed iron line profiles for four galactic sources XTE J1550-564, GRO J1655-40, Cyg X-1 and GRS
1915+105 obtained by Miller et al. (2004b). All of these data can be fitted by our model. 
In Figures 11 and 12 (the right hand panel) we show an example
of these fits to the data.   For GRO J1655-40 for the pure scattering model
 we find that optical depth $\tau=2.0$ and  an outflow velocity  is $ \beta = 0.1$,  much higher than that for MCG 6-30-15 and $E_{ph} = 7.1$ keV. 
If we take into account the photo-absorption along with scattering in our model, then the best-fit parameters are $\tau=3.3$, $\beta=0.25$ and  $E_{ph} = 7.1$ keV.
It is worth noting that the fit in this case is not so good as that in the case of the pure scattering model. It can mean that the iron abundance in GRO J1655-40 is less than cosmic
abundance. In fact, LT04 calculated the photoabsorption cross-sections for the cosmic abundances. 



As we have  pointed out, the MC simulated line profiles are  independent of the outflow size
and are determined by four  parameters $\tau$, $\beta$, $kT_e$ and $E_0$. The  size of the outflow
shell can be determined by timing characteristics such as variability timescales and time lags. 
Our model predicts soft time lags, i.e. the time lags become longer as energy decreases. 
Time lags are scaled with the light crossing time, $t_{cross}$, 
from one to  a few $t_{cross}$. Unfortunately  time lag information is not yet available  
for all  sources analyzed. The only available information at present is the lack of the iron line variability 
with respect to the continuum for a number of extragactic sources [Markovitz, Edelson \& Vaughan (2003)].
Thus one can suggest that  these line
features are formed in a more extended configuration than the source of the continuum (which is presumably a Compton cloud close 
to the compact object). Here we demonstrate how our model can be used in the  analysis of the line data.   
The detailed interpretation of the line data using our outflow model  will be presented in a separate publication.

\section{Outflow  reprocessing of X-ray continuum spectra: MC simulated spectra}
 We also  study the modification of the central source X-ray spectrum in the warm outflow as a result of photoionization and
 downscattering.  In Figure 13 we present the  simulations
 for different values of $\tau=2,~4$ and $\beta=0.1, ~0.3$. 
 The results of simulations are shown as $E^2 F(E)$ diagrams, where $F(E)$ is a photon spectrum.
 In these simulations, the incident (central source) spectrum is approximated by the
 sum of two components: a blackbody shape radiation presumably coming from the inner part of the accretion disk,
 and a power-law with an exponential cutoff coming from the Compton cloud (corona). 
  The parameters used for this input spectrum are the color temperature of the blackbody component $kT_{bb} = 1.2$ keV, 
  the power-law photon index  $\Gamma=1.5$ (energy index $\alpha=0.5$) , and the cutoff $E_{cutoff}= 50$  keV. As shown in Comptonization theory (see e.g. ST80, Titarchuk 1994) the exponential cutoff energy $E_{cutoff}$ of the unsaturated Comptonization spectrum
 is approximately $2kT_e$ where $kT_e$ is the plasma temperature of the Compton cloud.  The  temperature of spherically symmetric wind is  $kT_e = 0.1$ keV.
 
For the X-ray spectrum of the central source we assume that the photon numbers are the same for the blackbody and hard
components. In fact, the ratio of the photon numbers depends on the illumination of the Compton cloud by the source of blackbody radiation.


 The photoelectric effect is taken into account in these simulations.
When a photoelectric event occurred, we drew a random number to firstly determine if this event 
occurs in the K shell of iron. The probability 
of fluorescence is about 30 \%. 
The fluorescence photon is then  propagated into the cloud in the same way as the input central 
object spectrum. 

In the simulated spectra one can clearly see features of the strong fluorescent K$_{\alpha}$ line and the deep K-edge formed in the wind. 
The prominent bump around 25 keV is a combined effect of photoionization and downscattering in the wind. 
The wind modification of the spectra is very sensitive to the optical depth of the wind $\tau$ and $\beta$.
For example, for a given $\beta=0.1$ (or the wind velocity $0.1c$ ) the depth of K-edge is broader and much deeper for $\tau=4$ than  for $\tau=2$.
As expected the width of the line increases with $\beta$ (compare the spectrum for $\beta=0.1$ with that for $\beta=0.3$).
It is important to emphasize that the shape of K$_{\alpha}$  line consists of narrow and broad components. These narrow and broad line  features of the line 
are really  observed in Cyg X-1 (see review of the K$_\alpha$ line observations in Cyg X-1 in the introduction section of ST06). 
  
The shape of the spectrum for $\tau=4$ and $\beta=0.3$ is very similar to the observed spectrum of Cyg X-3 (TSh05 and Trudolyubov, 2006, private communication). 
We present the interpretation of Cyg X-3 data based on the results of our simulations elsewhere.

\section{ Discussion and Conclusions}

It is important to point out  that  Thomson optical depth of outflow is an order of unity 
when the mass outflow rate $\dot M_{out}$
is an order of the Eddington mass accretion rate $\dot M_{\rm Edd}$ and higher 
(see e.g. formula 4 in King \& Pounds 2003).

Recent {\it XMM-Newton} observations of bright quasars (Pounds et al., 2003a,b; Reeves et al., 2003) give strong evidence
for powerful outflows from the nucleus with mass rates $\dot M_{out}\sim \dot M_{\odot}$ yr$^{-1}\sim \dot M_{\rm Edd}$ and
velocity $V\sim0.1c$ (i.e. $\beta=0.1$) in the form of blueshifted X-ray absorption lines. These outflows closely resemble those 
recently inferred in a set of ultraluminous X-ray sources with extremely soft spectral components 
(Mukai et al. 2003; Fabbiano et al. 2003).
It is also worth noting the recent  observations of the strong outflows from GRO J1655-40 
(Miller et al. 2006) and the Nucleus of the Seyfert 1 Galaxy NGC 3783 using 
an intensive HST/Chandra/FUSE monitoring (Crenshaw et al. 2004) and the recent Chandra observations  of highly obscured AGN (Levenson et al. 2004) reported in the AAS HEAD meeting in  2004. In particular, Levenson et al. analyze new Chandra observations
 of three
Compton thick Seyfert 2s with obscuring column densities that exceed $10^{24}$ cm$^{-2}$. They report that the Fe K$_\alpha$ line is very
prominent in these examples with equivalent width $>1$ keV.
 Thus one can  conclude that powerful mass outflows from
Eddington-limited accreting compact objects appear to be a very widespread phenomenon. 
King \& Pounds (2003) and Begelman, King \& Pringle (2006)  argue that the powerful outflows 
 may provide the high luminosity observed in quasars and ULXs, and imply that such objects have a major effect on their surroundings.
They further suggest that the powerful outflow phenomena have barely been explored, and the field promises to be fruitful.
 
It is out of the scope of this paper to study the  consequences of the X-ray  radiation for hydrodynamics of these outflows.
Our goal is to elaborate the spectroscopic tools for probing the outflow physical conditions such as column density 
(the outflow optical depth) temperature and iron ionization charge state. Moreover the photoionization modeling of the outflow (see LT04) permits the
determination of its distance from the central black hole.

We have presented  the Monte Carlo simulations of the radiative transfer of
monochromatic photons and continuum photons within a diverging outflow, a problem which is also of observational 
interest, in view of the observations of broad, redshifted Fe lines and downscattering bumps (so called ``reflection features'') in  
extragalactic and galactic black hole  candidate spectra. 
We also provide an analytic description of the simulated spectra which allows us to understand, in  detail,
 the effect of  the dowscattering modification of the central source spectrum.
In the  form presented the derived formulae of the resulting spectra can be easily used  as an efficient  analytical tool for the spectral data analysis.
We demonstrate that the strength of the downscattering bump is determined by the average number of scattering that photons undergo in the wind. 
From our simulations of the pure scattering wind model we find that the average number of scattering $N_{av}=b^{-1}$ 
suffered by the photons in the wind  of optical depth $\tau\gax 2$ is $\gax \tau^2$ for $\beta\lax 0.1$ (see \S 3 for details).
On the other hand the average number of scatterings $N_{av}$ decreases when the velocity of the wind increases. 
A larger  fraction of the photons are rather carried out by the flow than scattered  there
when velocities are higher than $0.1c$.
 
The outflow leads to a  redshift of the photon energy  producing a broad, red 
component (see Fig. 1 for illustration of this redshift effect) not unlike those of the Fe K$\alpha$ lines observed. 
We demonstrate analytically and numerically  (using Monte Carlo simulations) that the mean photon energy change per scattering  in the outflow $<\Delta E>$ is the first order effect with respect of $V/c$, namely $ <\Delta E>\propto-V/c$.
We  further emphasize   
that the red-skewed line features appear   when the line photons are generated very close to the bottom
of the outflow due to the outflow illumination by the X-ray  hard radiation of the central source.
 

An important result of our simulations 
(see also TKB's results) is the natural suppression of a blue wing in the iron line  feature without 
the need to invoke any specific geometric arrangement for the emission.
The illumination of the outflow from inside by the X-ray radiation, is  natural  for a compact source
geometry, where the X-ray radiation  originates from the innermost part of the source 
(a Compton  cloud  along with a disk).
The powerful wind  presumably starts in  outskirts of the source (see Everett \& Ballantyne 2004 for the
wind model).
In  our model  the red-wing  photons have undergone numerous  scatterings 
in an (effectively) optically thick medium, with
an associated first order redshift in each  scattering. 
It is not surprising that  
the blue wing,  as a result of the second order $(v/c)^2$ effect,  is weak for $kT_e\lax$ keV
and $\beta=V/c\lax0.1$. 

{\it This red-skewed line  is a  natural consequence of the first order Doppler effect 
 in the presence of multiple scattering and photo-absorption events in the wind.
It does not require  any particularly fine tuned geometric arrangement}.

LT04 find the self-consistent temperature and ionization structure 
of the wind shell as a function of the parameter (radius/luminosity, $r_{in}\tau/L_{40}$ where $r_{in}$ is 
the inner outflow radius and $L_{40}$ is X-ray luminosity in units of $10^{40}$ erg s$^{-1}$) 
for a given Thomson optical depth of
the shell. It is evident that the ionization parameter $L_{40}/nor^2$ is constant  through the shell (i.e.
for $r>r_{in}$) if the velocity of the wind is
constant through the flow.  Thus  the LT04 solution allows one to determine the size of the shell base for a given
luminosity of which the red-skewed line is observed. LT04 derive the range of parameter $r_{in}/L_{40}$ 
where the outflow-produced  stable solution exist. The allowed $r_{in}/L_{40}$ values are concentrated around $10^{12}-10^{13}$ cm. 

{\it Our model predicts the soft lags to be a function of $E/E_0$, $\beta$ and $\tau$}.
The combination of the red-skewed features along with the soft time lags 
that strongly depends on $\beta$, $\tau$ and $kT_e$ are intrinsic signatures of any diverging outflow. 

The demonstrated application of our outflow model to  data   points out a potentially 
powerful spectral  diagnostic for probes of the outflow-central object connection in Galactic and 
extragalactic BH sources.  
Particularly, analyzing  MCG-6-30-15 data (see \S4) we infer that bulk outflow  velocity is about 6000  km s$^{-1}$.
On the other hand  the OVII edge in the warm absorber of MCG-6-30-15 indicates that the flow along line of sight is
about $5000\pm1000$ km s$^{-1}$ (Fabian et al. 1994). 
Thus  using a new X-ray spectroscopic method [our diverging (outflow) model]
  we confirm  the early finding (that used UV spectroscopy)  
  that velocity of the warm absorber outflow of MCG-6-30-15 is about 5000 km s$^{-1}$. 

  
L.T. acknowledges the support of this work by the Center for Earth Observing 
an Space Research of the George Mason University. L.T. also appreciates  productive  discussions 
with Ralph Fiorito, Stuart Wick, Phil Uttley, Martin Laming and Chris Shrader.
We also acknowledge discussion of the paper results with
the referee
and his/her constructive and interesting suggestions.  
\appendix

\section{Downscattering spectrum. Derivation of Analytical Spectra} 
\subsection{Power-law as a spectrum of the primary photons}

Here we demonstrate how  formula (\ref{emerg})  can be analytically calculated using the steepest descent method.
At first to illustrate the main idea of  this derivation we consider the simplest case
when $\varepsilon=0$ and $\varphi(z)=z^{-\alpha}$ when 
\begin{equation}
W^{(0)}(z,u)=(1/z-u)^{\alpha-1}
\label{w0_power}
\end{equation}
and integral (\ref{emerg}) is presented by  equation (\ref{powerlaw}), 
where ${\cal P}(u)=b\exp(-bu)$.
If we change the variable $v=(1/z-u)z$ under integral (\ref{powerlaw}) then the integral can be rewritten as follows:
\begin{equation}
{\cal F}_E(z)= \frac{Abz^{-\alpha}}{z}\exp(-b/z)\int_{0}^{1}v^{\alpha-1}\exp(bv/z)dv.
\label{powerlaw_ap1}
\end{equation}
Because the dimensionless frequency $z=E/m_ec^2\ll1$ for the energy range of the interest ($E\ll 511$ keV) and $b\gax0.1$ 
there is a wide energy range where $\lambda=b/z$ is a big parameter. Then the integral in formula (\ref{powerlaw_ap1}) can be
calculated analytically using the steepest descent method. In order to do it one should expand the integrand function 
$v^{\alpha-1}$ in the Taylor's series 
\begin{equation}
v^{\alpha-1}=1+\frac{(1-\alpha)}{1!}(1-v)+\frac{(1-\alpha)(2-\alpha)}{2!}(1-v)^2+...
+\frac{(1-\alpha)(2-\alpha)...(n-\alpha)}{n!}(1-v)^n +...
\label{v_expansion}
\end{equation}
near the point $v=1$ {\it where $\exp(bv/z)$ has a sharp maximum}. 
It is worth noting that  series (\ref{v_expansion}) converges for all $v<1$ and thus
the series presentation of the emergent spectrum (\ref{powerlaw_ap1}) is {\it exact} for any $\lambda=b/z$:
$$
{\cal F}_E(z)= \frac{Abz^{-\alpha}}{z}\exp(-b/z)[L_{0}(\lambda)+
\frac{(1-\alpha)}{1!}L_{1}(\lambda)+\frac{(1-\alpha)(2-\alpha)}{2!} L_{2}(\lambda)+...
$$
\begin{equation}
+\frac{(1-\alpha)(2-\alpha)...(n-\alpha)}{n!}L_{n}(\lambda) +...],
\label{powerlaw_ap2}
\end{equation}
 where 
\begin{equation}
 L_{n}(\lambda)=\int_0^1(1-v)^n\exp(\lambda v)dv~~~~~{\rm~for}~~n=0,~1,~2~... 
\label{L_func}
 \end{equation}
If we  introduce the new variable $u=(1-v)/z$ in integral (\ref{L_func}) then  formula (\ref{powerlaw_ap2}) can be rewritten
in the form 
$$
{\cal F}_E(z)= Az^{-\alpha}[ M_{0}(z^{-1})+
\frac{(1-\alpha)}{1!}z M_{1}(z^{-1})+\frac{(1-\alpha)(2-\alpha)}{2!}z^2 M_{2}(z^{-1})+...
$$
\begin{equation}
+\frac{(1-\alpha)(2-\alpha)...(n-\alpha)}{n!} z^{n}M_{n}(z^{-1}) +...],
\label{powerlaw_ap3}
\end{equation}
 where 
$M_{n}(z^{-1})$ is defined by equation (\ref{M_func}).
\subsection{Power-law with exponential cutoff as a spectrum of primary photons}
In the general case of the incident spectra and $\varepsilon>0$ formula (\ref{emerg}) for the emergent spectrum can be rewritten as
\begin{equation}
{\cal F}_\nu(z,\varepsilon)=\frac{b}{z}
\int_0^{1}W^{(\varepsilon)}[z,vu_{max, \varepsilon}(z)] \exp(-\lambda v)dv 
\label{emerg1}
\end{equation}
where $\lambda=bu_{max, \varepsilon}(z)\gax 1$ and 
$W^{(\varepsilon)}(z,u)=e^{-4\varepsilon u/3}[\psi^{(\varepsilon)} (z,u)]\varphi[\psi^{(\varepsilon)}(z,u)]$ and $u=vu_{max, \varepsilon}(z)$. 
Because the integrand has a sharp exponential maximum at $v=0$ (or $u=0$) one can expand the non-exponential function  $W^{(\varepsilon)}(z,u)$
in series over $v$ (or $u$). That leads to the following formula
$$
{\cal F}_{\nu}(z,\varepsilon)= \frac{A}{z}[W^{(\varepsilon)}(z,0)M_0(u_{max, \varepsilon})
+W^{(\varepsilon)(1)}_{u}(z,0)
M_1(u_{max, \varepsilon})/1!+
$$
\begin{equation}
+W^{(\varepsilon)(2)}_{u}(z,0)M_2(u_{max, \varepsilon})/2!+...
+W^{(\varepsilon)(n)}_{u}(z,0)M_n(u_{max, \varepsilon})/n!+...],
 \label{intexpan1}
\end{equation}
where $W^{(\varepsilon)(n)}_{u}(z,u)$ is $n$th partial derivative of  $W^{(\varepsilon)}(z,u)$ over $u$
and the moments $M_n(x)$ ($x=u_{max, \varepsilon}$)
of ${\cal P}(u)$ as follows 
\begin{equation}
M_n(x)=-x^n\exp(-x)+nM_{n-1}(x)/b~~~~~{\rm~for}~~n=1,~2~...
\label{M_n}
\end{equation}
For example for n=0,~1,~2 we obtain
\begin{equation}
M_{0}(x)=1-\exp(-bx),
\label{moment0}
\end{equation}
\begin{equation}
M_{1}(x)=-x\exp(-bx)+(1/b)M_0(x),
\label{moment1}
\end{equation}
\begin{equation}
M_{2}(x)=-x^2\exp(-bx)+(2/b)M_1(x).
\label{moment2}
\end{equation}
Because 
\begin{equation}
W^{(\varepsilon)(n)}_{u}(z,0)=W^{(\varepsilon)}_{u}(z,0)\prod^{n-1}_{i=0}[\ln W^{(\varepsilon)(i)}_{u}(z,u)]_{u}^{\prime}(u=0)
\label{w__der}
\end{equation}
and 
\begin{equation}
W^{(\varepsilon)}(z,0)=z\varphi(z) 
\label{w__0}
\end{equation}
the analytical formula (\ref{intexpan1}) for the emergent spectrum  can be presented as
\begin{equation}
{\cal F}_{\nu}(z,\varepsilon)= 
\varphi(z)\left\{M_0(u_{max, \varepsilon})+\sum_{n=1}^{\infty}
\left[\prod^{n-1}_{i=0}[\ln W^{(\varepsilon)(i)}_{u}(z,u)]_{u}^{\prime}(u=0)\right]
M_n(u_{max, \varepsilon})/n!\right\}.
\label{intexpan3}
\end{equation}
Thus the resulting spectrum ${\cal F}_{\nu}(z,\varepsilon)$ is a product of the incident spectrum $\varphi(z)$
and the downscattering transformation function (expression in the curl parenthesis in Eq. \ref{intexpan3}) 
 that depends on the properties of the outflow, velocity, optical depth via the parameter $b$.

In fact, the more convenient way to calculate the resulting spectrum ${\cal F}_{\nu}(z,\varepsilon)$  is to calculate 
the derivatives $W^{(\varepsilon)(n)}_{u}(z,u)$  explicitly   using the factorized form of $W^{(\varepsilon)(1)}_{u}(z,u)$, namely
\begin{equation}
W^{(\varepsilon)(1)}_{u}(z,u)=W^{(\varepsilon)}_{u}(z,u)\Phi(z,u),
\label{firsderiv}
\end{equation}
where 
\begin{equation}
\Phi(z,u)=[\ln z_0\varphi(z_0)]_u
\label{phi_z_def}
\end{equation}
and 
\begin{equation}
\varphi(z_0)=z_0^{-\alpha}\exp(-z_0/z_\ast).
\label{pl_expcutoff}
\end{equation}
Using Eqs. (\ref{phi_z_def} , \ref{pl_expcutoff})  we obtain a formula for  $\Phi(z,u)$: 
\begin{equation}
\Phi(z,u)=[\ln z_0\varphi(z_0)]_u=-4\varepsilon/3+\frac{1-\alpha}{z_0}\frac {\partial z_0}{\partial u}-\frac{1}{z_\ast}
\frac{\partial z_0}{\partial u}. 
\label{phi_z}
\end{equation}
Then one can calculate $W^{(\varepsilon)(n)}_{u}(z,u)$ as follows
\begin{equation}
W^{(n)}_{u}= \sum_{i=0}^{n-1}C_{n-1}^i W^{(n-1-i)}_u\Phi^{(i)}_u~~~ {\rm for}~~~n=2,3,...  
\label{nder}
\end{equation}
where $C_n^k=n(n-1)..(n-k-1)/k!$ and $ W^{(0)}_u=W$. Here and below we omit superscript  $\varepsilon$ in order to 
 simplify the notation for  $W^{(\varepsilon)(n)}_{u}(z,u)$ and its derivatives. We also introduce new notations $W^{(i)}_u=W^{(i)}(z,u)_u$ and 
 $\Phi^{(i)}_u=\Phi^{(i)}(z,u)_u$.

Because 
\begin{equation}
\frac{\partial z_0}{\partial u}=(\varepsilon/3)z_0+z_0^2
\label{z_0vsdu}
\end{equation}
we can write Eq.(\ref{phi_z}) as
\begin{equation}
\Phi(z,u)=-4\varepsilon/3+(1-\alpha)(\varepsilon/3+z_0)-\frac{1}{z_\ast}
\frac{\partial z_0}{\partial u} 
\label{phi_z1}
\end{equation}
and find the derivatives of $\Phi(z,u)$ as 
\begin{equation}
\Phi^{(n)}_u(z,u)=(1-\alpha)\frac{\partial^{n} z_0}{\partial u^n}-\frac{1}{z_\ast}\frac{\partial^{n+1} z_0}{\partial
u^{n+1}}~~~{\rm for}~~n=1,2,3... 
\label{phi_z_der}
\end{equation}
The second and higher derivatives of $z_0$ over $u$ are
\begin{equation}
\frac{\partial^2 z_0}{\partial u^2}=(\varepsilon/3)\frac {\partial z_0}{\partial u}+2z_0\frac {\partial z_0}{\partial u},
\label{2z_0vsdu}
\end{equation}
\begin{equation}
\frac{\partial^n z_0}{\partial u^n}=(\varepsilon/3)\frac{\partial^{n-1}z_0}{\partial u^{n-1}}+
+2\sum_{i=0}^{n-2}C_{n-2}^{i}\frac{\partial^i z_0}{\partial u^i}
\frac{\partial^{n-1-i} z_0}{\partial u^{n-1-i}}~~~{\rm for}~~~n=3,4...
\label{highz_0vsdu}
\end{equation}
Now all derivatives $W^{(n)}_{u}$  in Eq.(\ref{nder}) can be calculated very easily because $z_0=z$ at $u=0$ by definition.
\newpage
\section{The mean photon energy change in the bulk outflow. Analytical evaluations}
\subsection{Diffusion  method}
There is a standard procedure to estimate the energy gain and loss per scattering using diffusion approximation 
(Prasad et al. 1988 and  Titarchuk, Mastihiadis \& Kylafis 1997, Appendix D). 
Let us assume that in  the initial moment $t=0$ the monochromatic line of energy $E_0$ emits   at any point of the scattering medium (outflow).  The  energy change of the line  $<\Delta E> $ due to scattering can be found using      
the time-dependent diffusion kinetic equation:
\begin{equation}
\frac{\partial n}{\partial u}=L_r n +L_{\nu} n
\label{kinet_eq}
\end{equation}
where $n(r, \nu)$ is the photon occupation number,  $L_r$, $L_{\nu}$ the space and  energy operators respectively
(see TKB, Eq. 3) and $u=\kappa ct$ is the time measured in units of mean time between scatterings
(see Rybicki \& Lightman 1979, ST80).
Our goal is to find a small change of the  radiation field background which at the initial moment   is 
 \begin{equation}
z^3n|_{u=0}=z^3n_{0}=z\delta(z-z_0).
\label{background}
\end{equation}
The energy change per scattering $<\Delta z>=<\Delta E> /m_ec^2$ can be found  by multiplication 
of the kinetic equation  (\ref{kinet_eq})  by  $z^3$,  substitution of equation (\ref{background}) in the right hand side 
 and integration over $z$.
Thus we have
\begin{equation}
\frac{ \delta z}{\delta u} =<\Delta z>=\int _0^{\infty}\frac{1}{3}\frac{\nabla ({\bf V}/c)}{\kappa}z^4\frac{\partial n}{\partial z}dz,
\label{energy_change_kineq}
\end{equation} 
where $\delta z/\delta u$ is a photon energy change per scattering.

Integration by parts gives
\begin{equation}
<\Delta z>=-\frac{4}{3}\frac{\nabla ({\bf V}/c)}{\kappa}\int _0^{\infty}z^3 n_0dz.
\label{final_change}
\end{equation}
Substituting $z^3n_{0}=z\delta(z-z_0)$ (see Eq. (\ref{background}) in this equation, we obtain 
\begin{equation}
<\Delta E>= -\frac{4}{3}\frac{\nabla ({\bf V}/c)}{\kappa}E_0.
\label{deltaE_divergence_appen}
\end{equation}
Because 
\begin{equation}
\nabla ({\bf V}/c)=\frac{2\beta}{r}
\label{diverg}
\end{equation}
and 
$\kappa r=\sigma_{\rm T}n_e r\approx\tau$ for the electron number density of the constant velocity wind
we can rewrite equation (\ref{deltaE_divergence_appen}) in the form
\begin{equation}
<\Delta E>\approx - \frac{8}{3}\frac{\beta}{\tau}E_0. 
\label{deltaE_tau_beta}
\end{equation}

\subsection{Geometry of  photon scattering in the outflow }
In this section  we  evaluate $<\Delta E>$ using  equation  (\ref{energ_change})  and geometry of photon scattering in the outflow.   Ultimately, we should calculate the mean of the scalar product of 
$({\bf V}_1-{\bf V}_2)\cdot{\bf n}/c$ (${\bf n}$ is a unit vector along the
path of the photon scattered at the next point, see Fig. 1).
 
The length of the vector  
\begin{equation}
|\Delta {\bf V}|= |{\bf V}_1-{\bf V}_2|= 2V \sin{\theta/2}\approx v\sin{\theta}~~~{\rm for~~~\theta<1}
\label{delta_v}
\end{equation}
where $\theta$ is an angle between ${\bf V_1}$ and  ${\bf V_2}$.
Let $l$ is the mean free path  between two consecutive scatterings at $r_1$ and $r_2$ (see Fig. 1) then 
\begin{equation}
\cos{\theta}=\frac{r_1^2+r_2^2-l^2}{2r_1r_2}.
\label{cos_theta}
\end{equation} 
If $|r_1-r_2|/r_1=\Delta r/r<1$ ($r_1\approx r_2=r$) and $\Delta r <l^2/2r$ then we can estimate $1-\cos{\theta}$ as follows
\begin{equation}
1-\cos{\theta}\approx (l/r)^2/2,
\label{1_cos_theta}
\end{equation}  
 Thus we can find that
\begin{equation}
\sin{\theta}\approx l/r=1/(N_e\sigma_{\rm T}r)=r/(r_{in}\tau)\sim f/\tau
\label{l_r}
\end{equation}
when $l/r<1$ and where  the numerical factor $f\gax1$. 
The precise value of the factor $f$ can be evaluated using our MC simulations.
To derive  formula (\ref{l_r}) we use the electron number distribution of the constant velocity wind (see formula \ref{electrden}). 

The cosine of the angle $\gamma$ between $\Delta {\bf V}=({\bf V}_1-{\bf V}_2)$ and ${\bf n}$ is  
\begin{equation}
\cos\gamma=\cos\psi\cos(\pi-\theta/2)+\sin\psi\sin(\pi-\theta/2)\cos(\varphi-\varphi_0)
\label{gamma}
\end{equation}
where $\psi$ is a zenith angle between vector ${\bf n}$ and the normal to vector ${\bf V_1}$ directed toward vector ${\bf V_2}$ and $\varphi$ and $\varphi_0$ are meridian (longitudinal) angles of ${\bf n}$ and $\Delta V$ respectively.
Then the mean value of $\cos\gamma$ over half of sphere located to the left from the vector ${\bf V_1}$  is calculated as follows
\begin{equation}
<\cos\gamma>=\frac{1}{2\pi}\int_0^{2\pi}d\varphi\int_0^{\pi/2}\cos\gamma \sin\psi d\psi=-\frac{1}{2}\cos(\theta/2). 
\label{gamma_mean}
\end{equation}

Thus using Eq.(\ref{energ_change})  and Eqs. (\ref{delta_v}-\ref{gamma_mean}) 
one can relate the  mean energy change per scattering, $<\Delta E>$, for photons undergoing quite 
a few scatterings in the flow with $\beta$ as follows:  
\begin{equation}
<\Delta E>\approx -\beta f/(2\tau)\sqrt{1-(f/2\tau)^2} E_0.
\label{deltaE_app}
\end{equation}


\begin{figure}
\includegraphics[width=6.in,height=4.2in,angle=0]{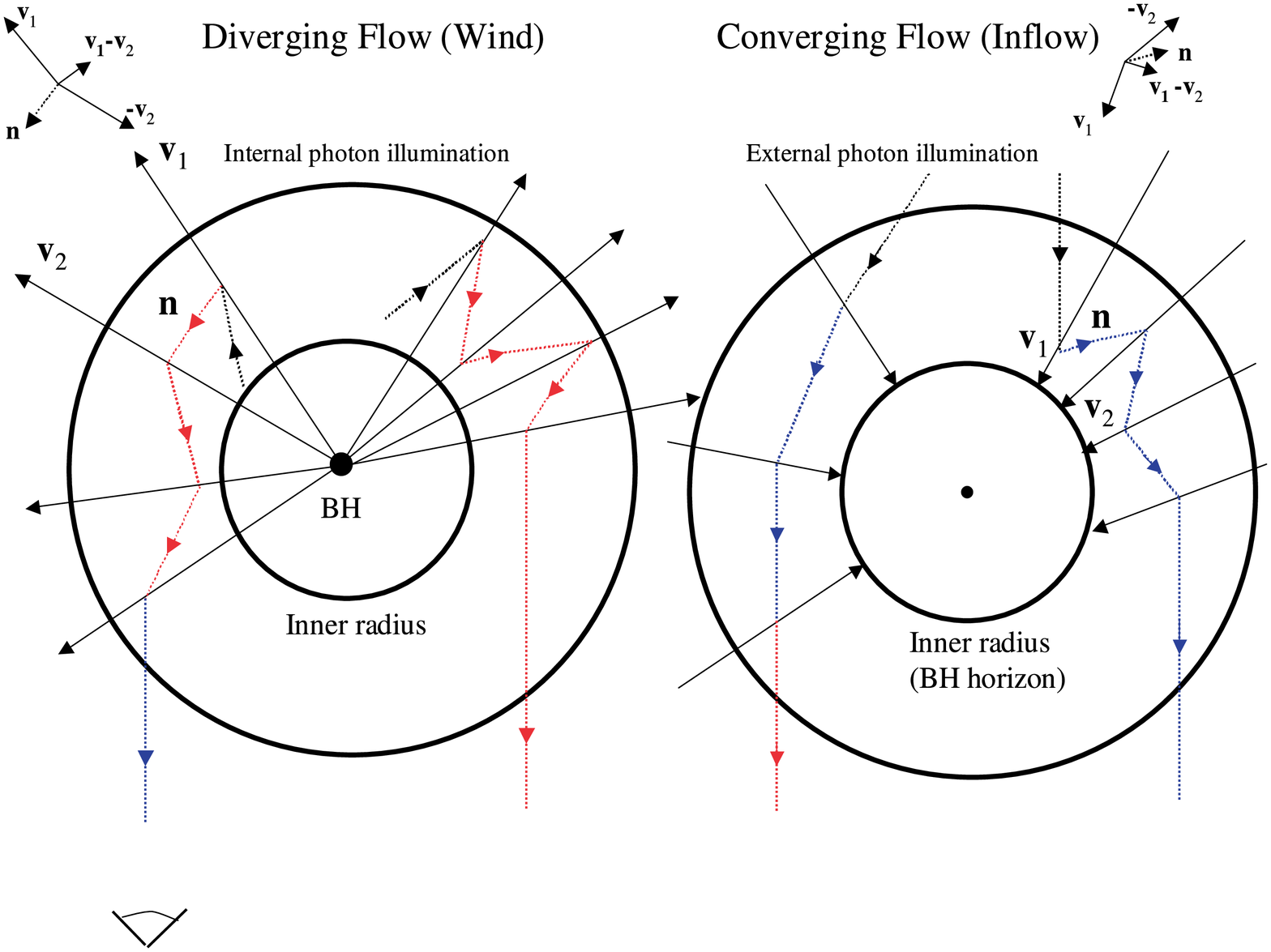}
\caption{On the left side: Schematic diagram showing wind geometry. The relatively cold  outflow (wind)
originates at the inner radius. The optical depth of the wind in
the Fe K continuum and the electron scattering optical depth of the wind  are
of order unity. A photon emitted near the inner boundary and subsequently scattered
 by an electron moving with velocity ${\bf V}_1$, is
received by an electron moving with velocity ${\bf V}_2$ as shown.  The change in frequency is
 $\nu _2 = \nu _1\left(1+\left({\bf V}_1-{\bf V}_2\right)
\cdot{\bf n}/c\right)$ where ${\bf n}$ is a unit vector along the
path of the mv photon scattered at the next point. In a diverging flow
$\left({\bf V}_1-{\bf V}_2\right)\cdot{\bf n}/c <0$ and photons are
successively redshifted, until scattered to an observer at infinity.
The color of photon path indicates the frequency shift in the rest frame of the receiver
(electron or the Earth observer).
On the right side: In a converging flow
$\left({\bf V}_1-{\bf V}_2\right)\cdot{\bf n}/c >0$ and photons are
blueshifted.
}
\end{figure}
\newpage
\begin{figure}
\includegraphics[width=6.in,height=6.5in,angle=0]{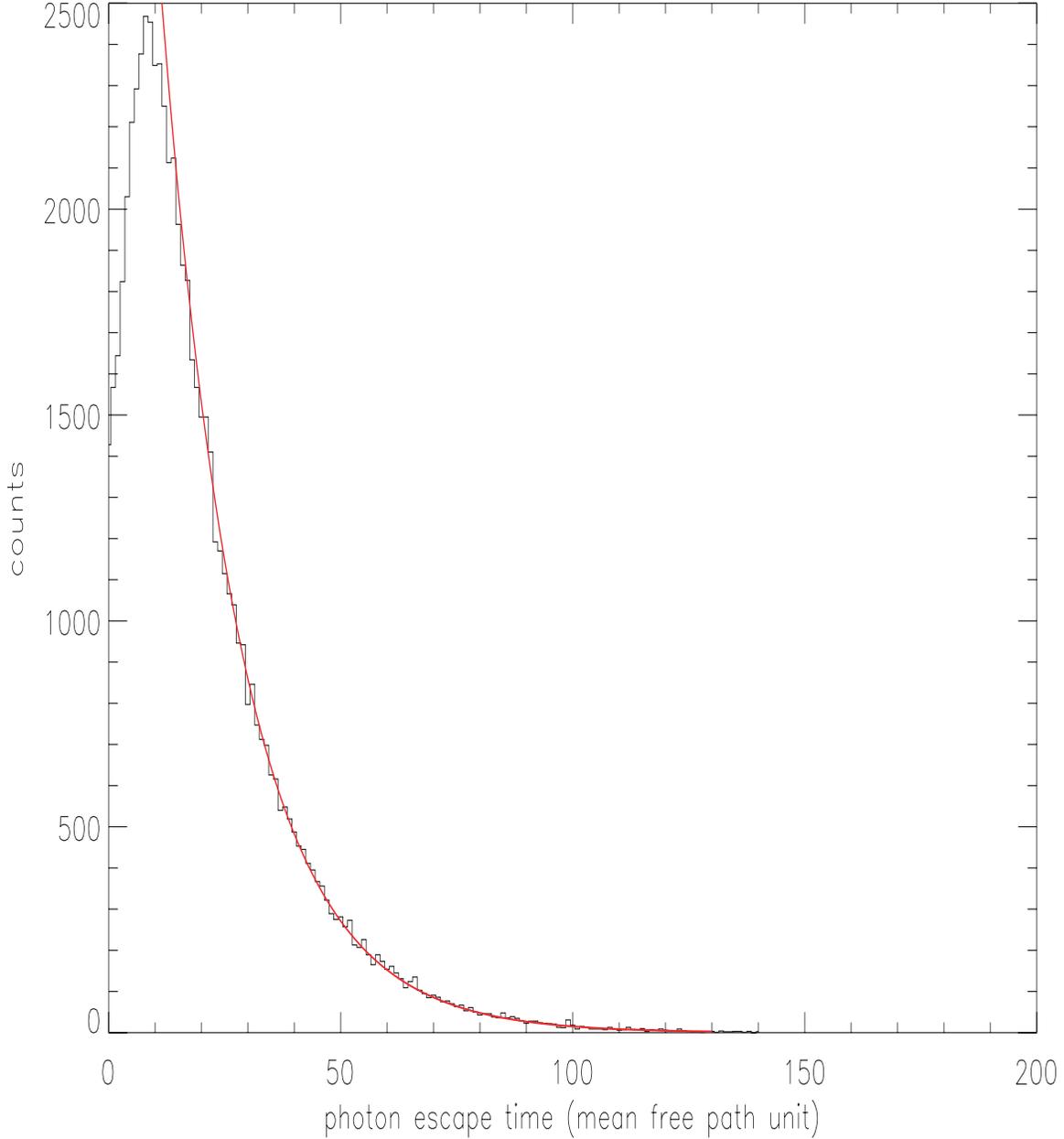}
\caption{Timing properties of the  outflow. The photon distribution over escape time 
${\cal P}(t/t_{fp})$, where $t_{fp}=l_/c$ is a free path time. This simulation is made for 
the incident spectrum as a power-law with exponential cutoff for $kT_e=0.1$ keV, $\tau = 4$ and $\beta =0.1$.
It is worth noting that the index $b=0.057$ of the exponential distribution $\exp(-bt/t_{fp})$ (red line) 
is very close to the best-fit parameter values $0.06\pm 0.005$ for the energy spectrum.  We find that in all cases the best-fit parameter $b$ obtained
from the photon distribution over escape time is very close to b obtained from the best-fit analytical spectrum.
}
\end{figure}

\begin{figure}
\includegraphics[width=4.9in,height=2.6in,angle=0]{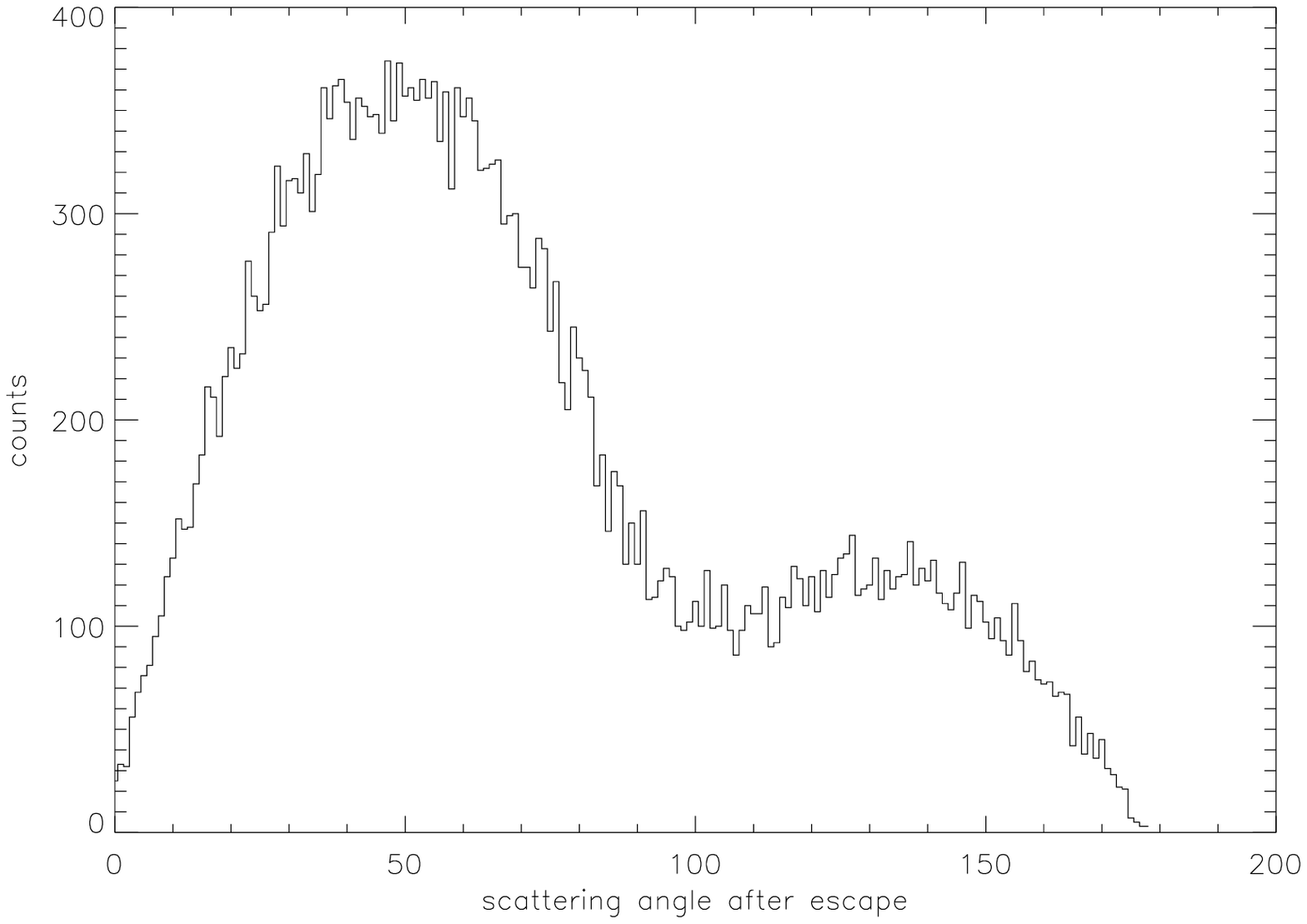}
\includegraphics[width=4.9in,height=2.6in,angle=0]{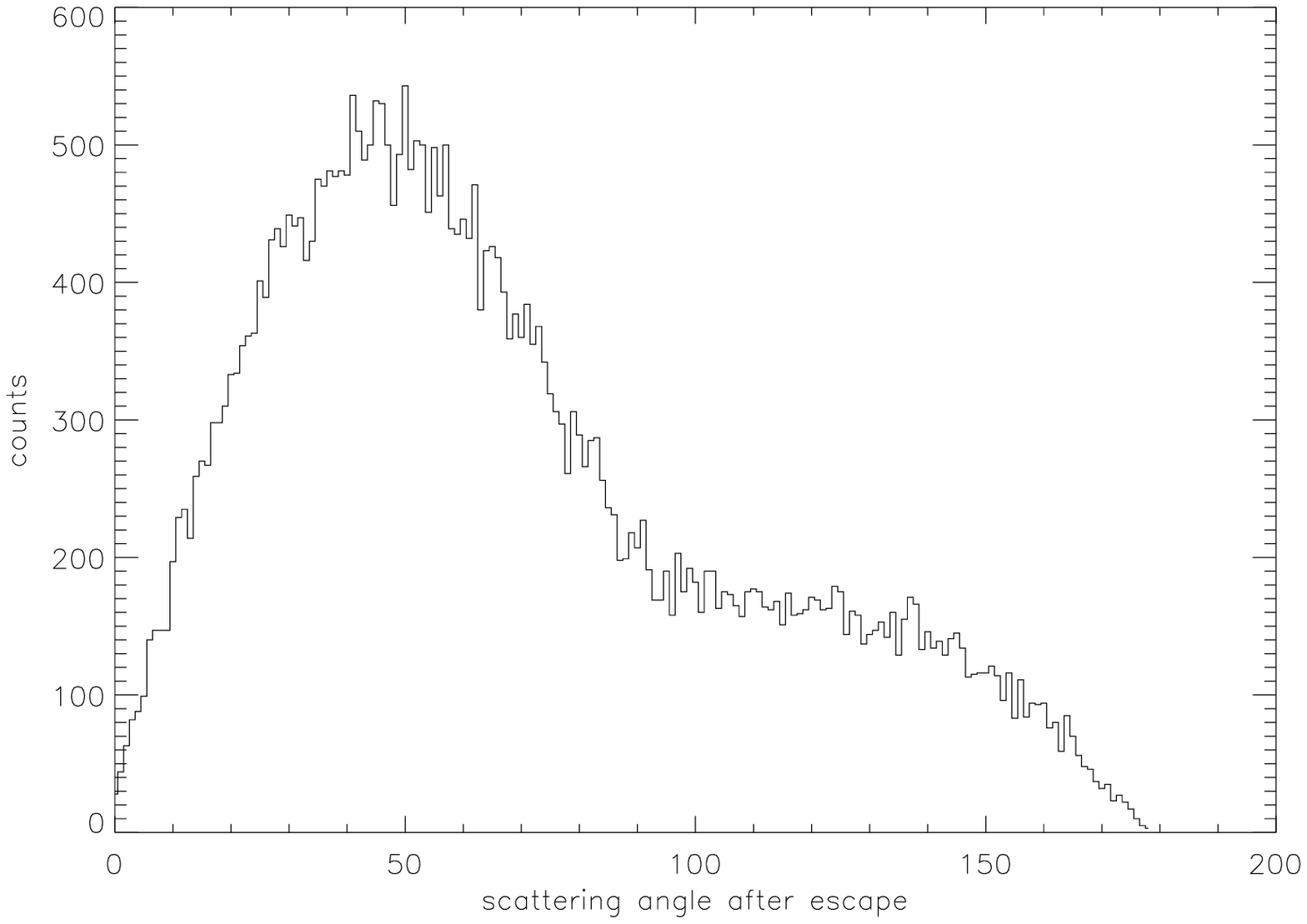}
\includegraphics[width=4.9in,height=2.6in,angle=0]{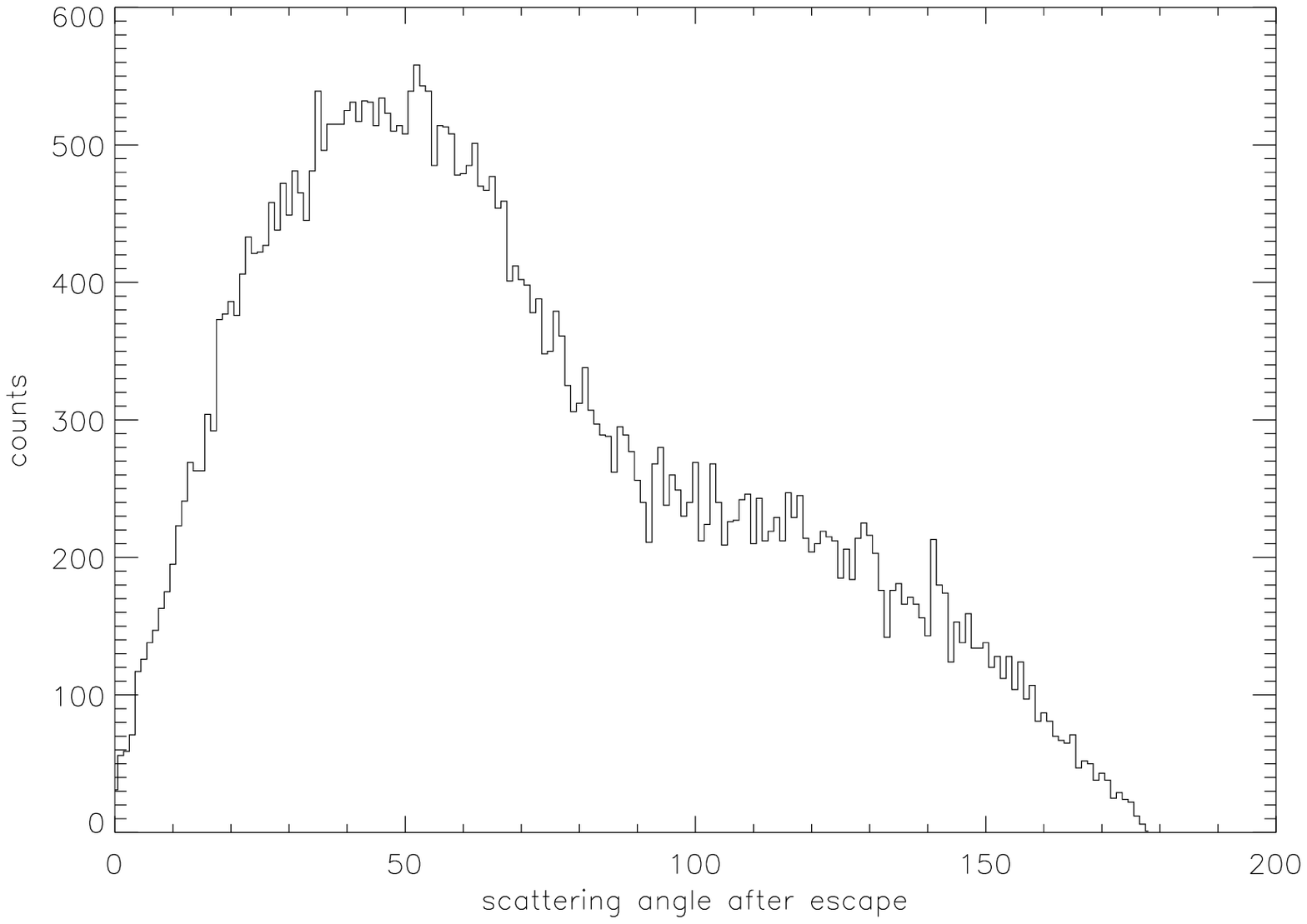}
\caption{The outflow shell is radially illuminated be the central source radiation. 
Photon distribution over deflecting angles with respect to the  radial direction of the incident photons.
Upper panel: $\tau=1$, middle panel: $\tau=2$ and lower panel: $\tau=4$ and  $\beta=0.1$. 
In fact, we find in our simulations that  the angle distribution of the scattering component   weakly depends on $\tau$. 
But there is a big difference in  the normalization  of the direct (non-scattering) component $A_N$ for which 
the ratio of $A_N(\tau=4)/A_N(\tau)=\exp(-4+\tau)$. 
}
\end{figure}

\newpage

\begin{figure}
\includegraphics[width=3.5in,height=3in]{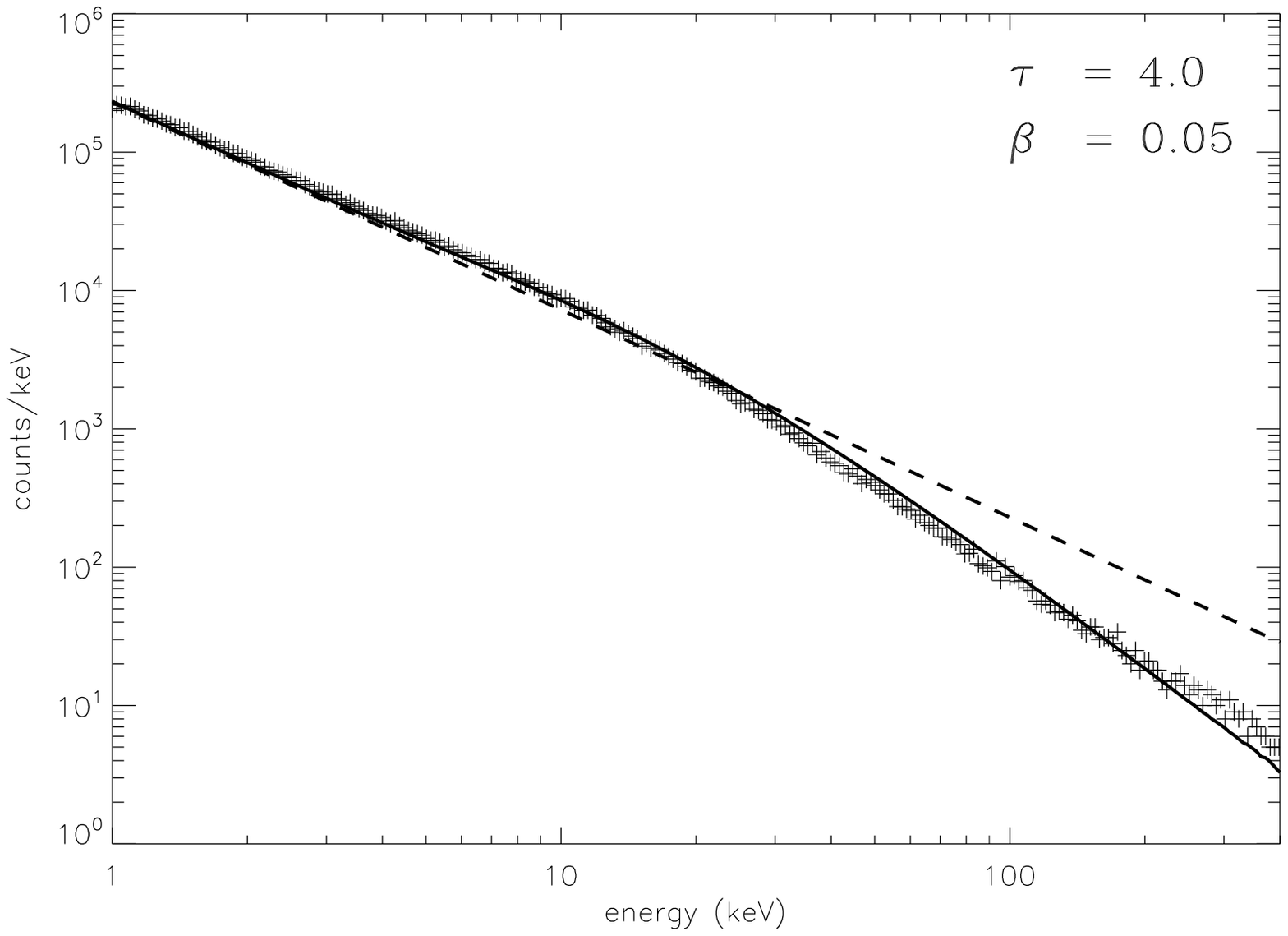}
\includegraphics[width=3.5in,height=3in]{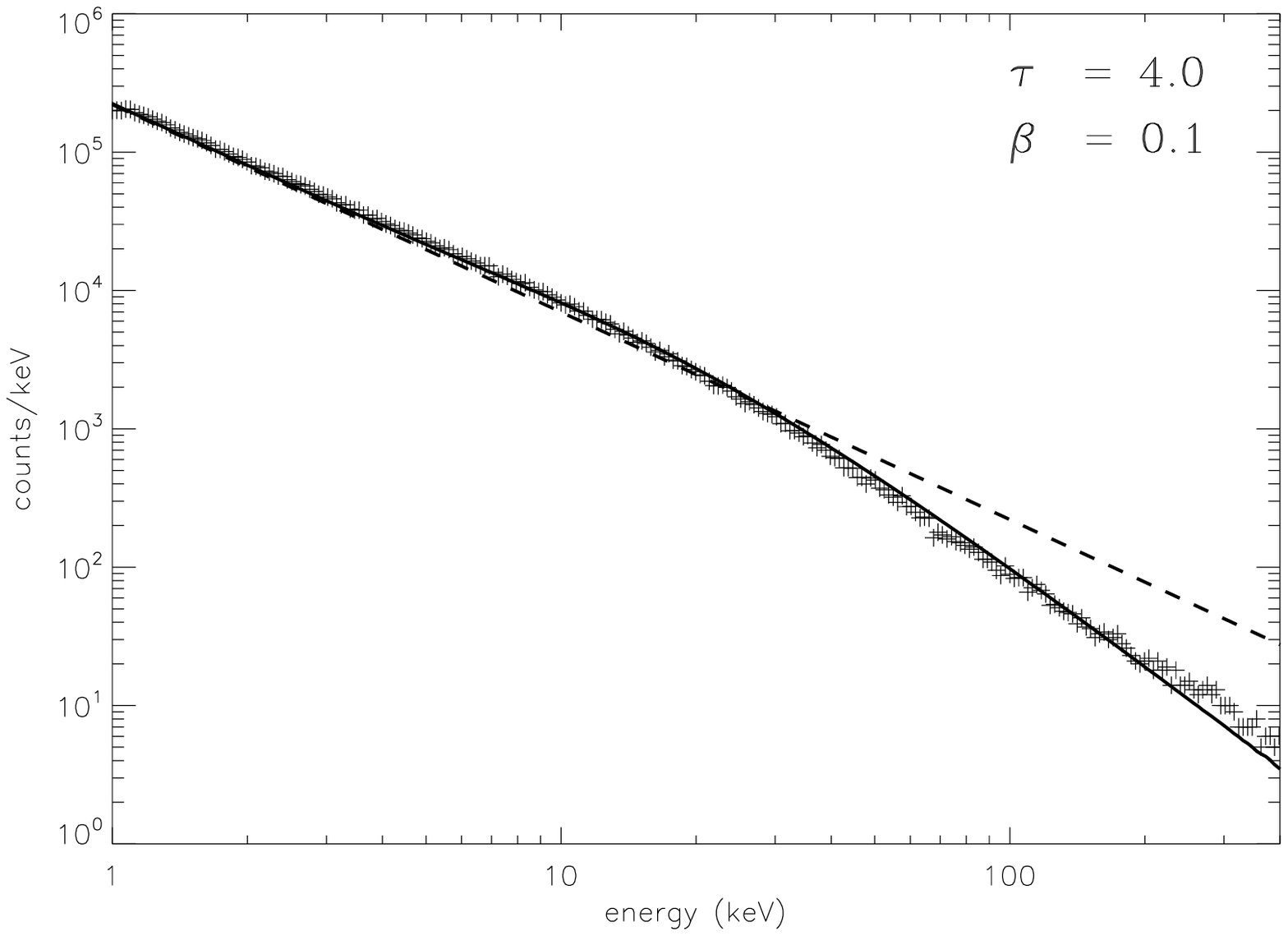}
\includegraphics[width=3.5in,height=3in]{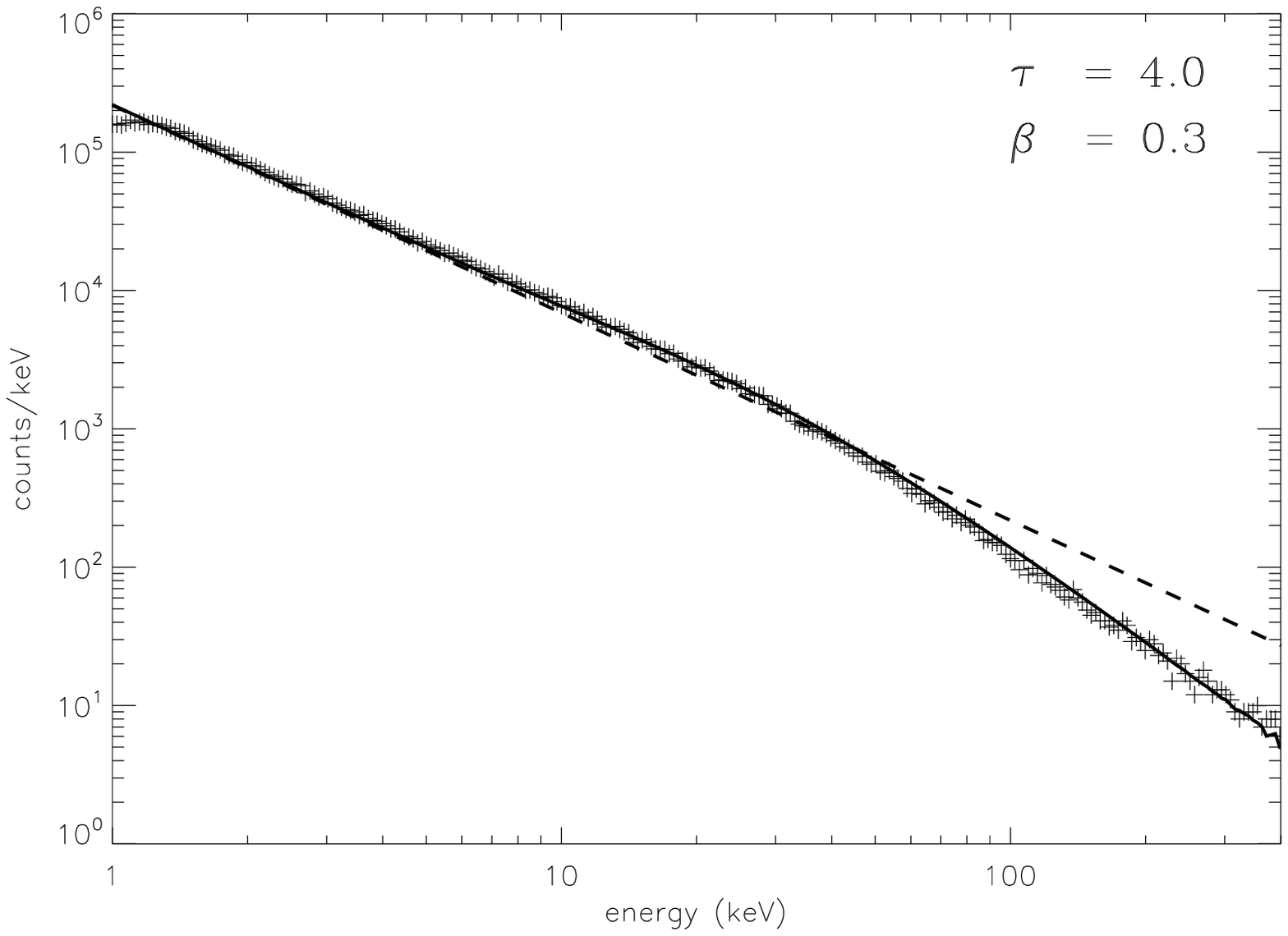}
\includegraphics[width=3.5in,height=3in]{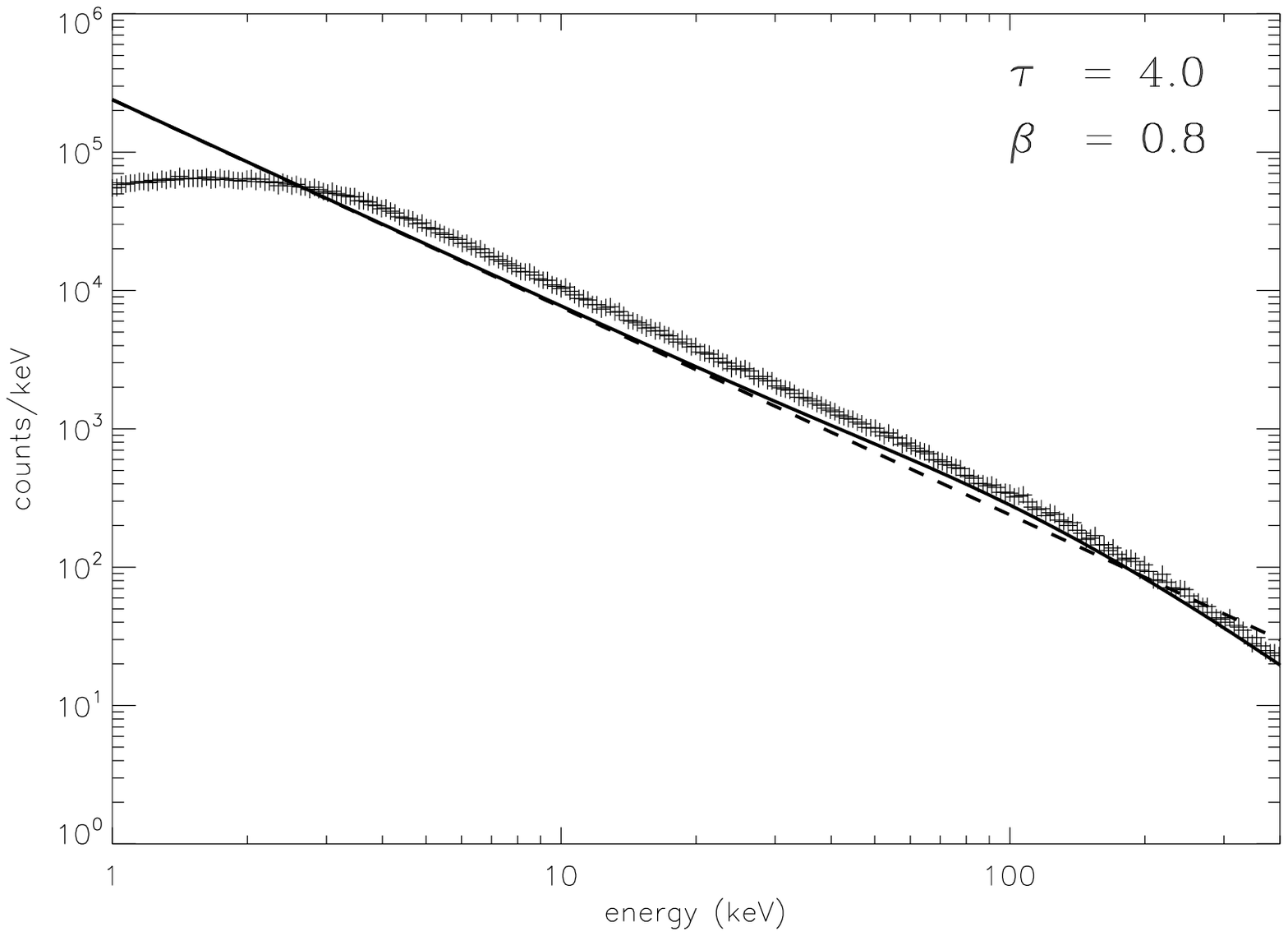}
\caption{Emerging spectrum of a compact object through an outflowing plasma. 
The incident spectrum  inside the outflowing plasma is taken to be a power-law
$\varphi(E) \propto E^{-\alpha}$  for which the energy index $\alpha=0.5$.
Outflow parameters are $kT_e=0.1$ keV, $\tau = 4$ and $\beta = 0.05,~0.1,~0.3,~0.8$.
The photon emerging spectrum (cross points) is  compared to the incident spectrum (dash line) and the analytical downscattering spectrum (solid line)
(see Eq. \ref{plexpan}).  For $\beta = 0.05,~0.1,~0.3,~0.8$, the best-fit parameters are $b=0.07,~0.07,~0.11,~0.45$ respectively.
  For all cases we fix the $\varepsilon=0$. 
One can clearly see a  bump in the MC simulated and analytical spectra, in the low energy band around 10 keV for all  $\beta\lax0.3$. 
Indeed, in the purely diffusive case that is shown  in this plot, the photon number is conserved, 
so the downscattered high energy photons, which are removed from the high energy part of the incident spectrum, are detected at lower energy. 
}
\end{figure}
\newpage
\begin{figure}
\includegraphics[width=3.5in,height=3in]{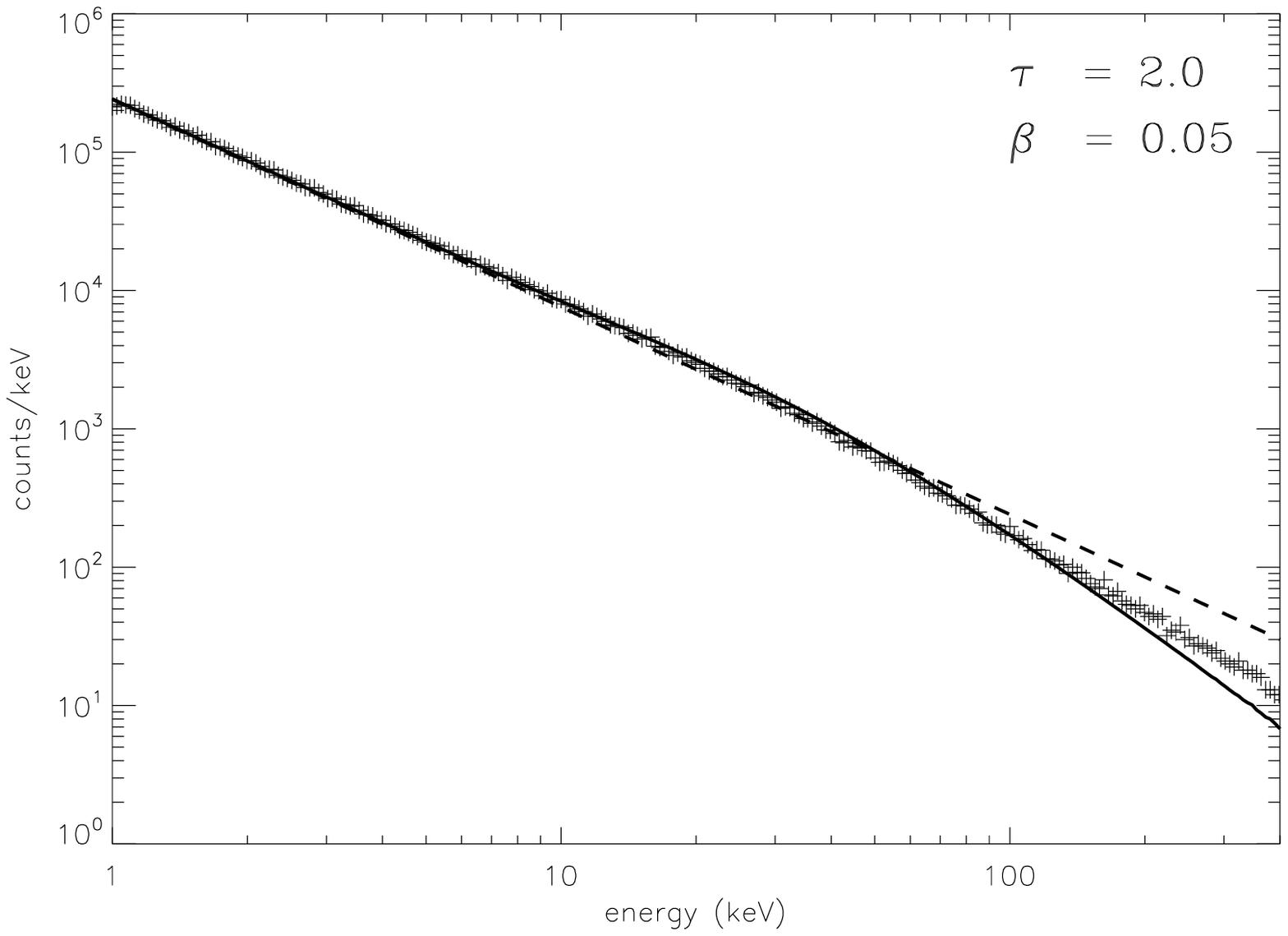}
\includegraphics[width=3.5in,height=3in]{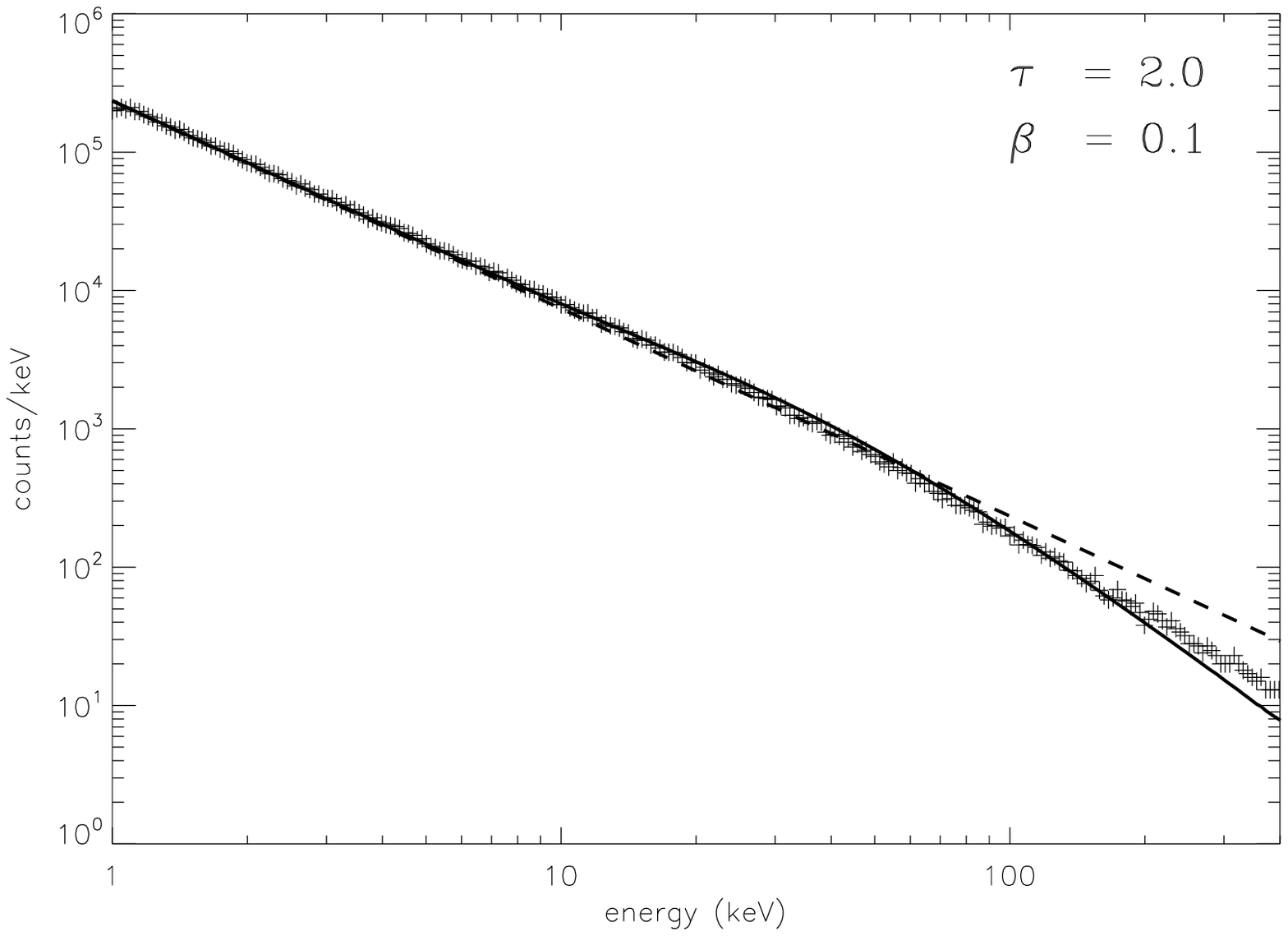}
\includegraphics[width=3.5in,height=3in]{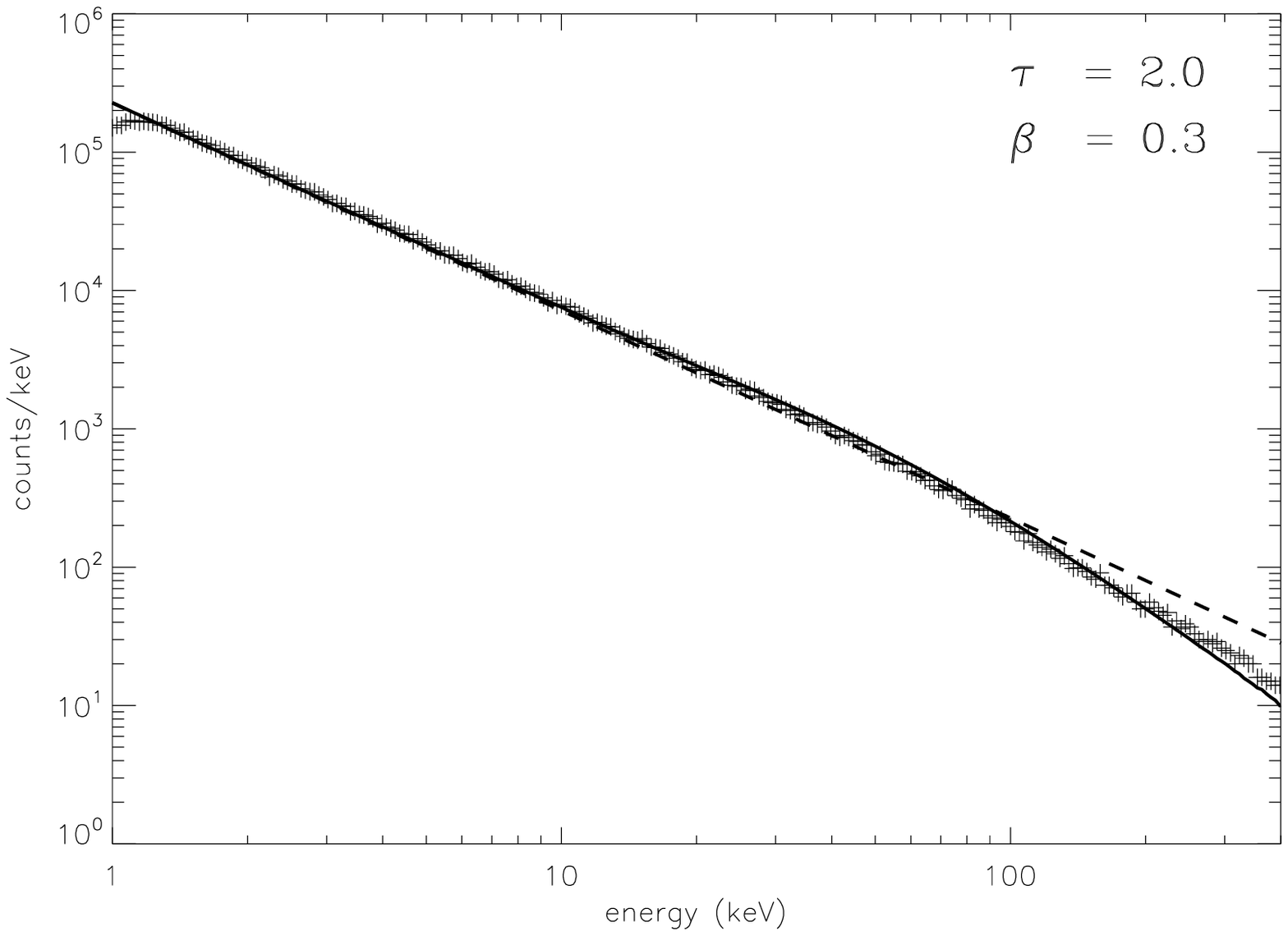}
\includegraphics[width=3.5in,height=3in]{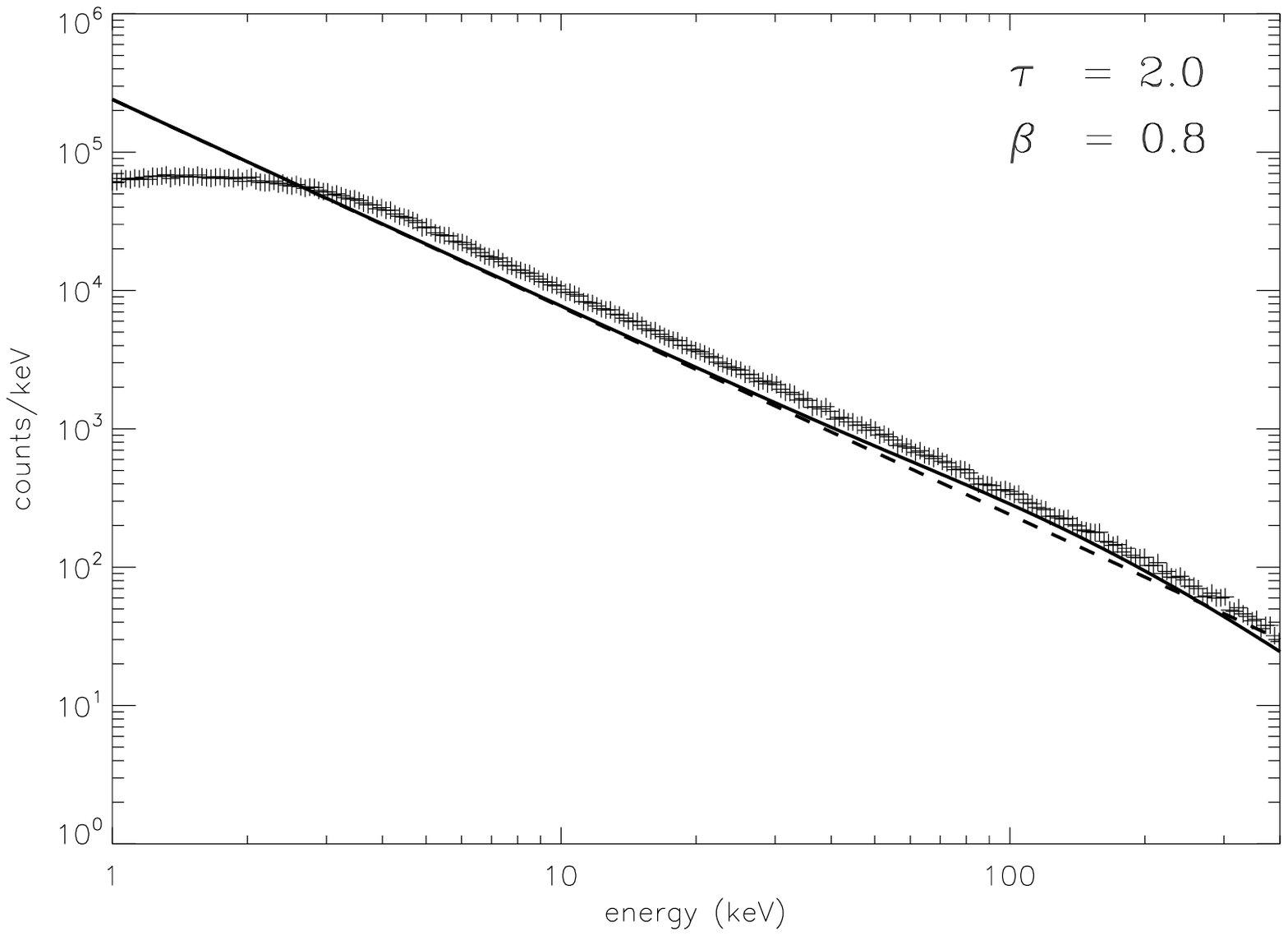}
\caption{ The same as in Fig. 4 for $\tau=2$. For $\beta = 0.05,~0.1,~0.3,~0.8$, the best-fit parameters 
are $b=0.13,~0.15,~0.21,~0.62$ respectively. 
}
\end{figure}

\newpage

\begin{figure}
\includegraphics[width=3.5in,height=3in]{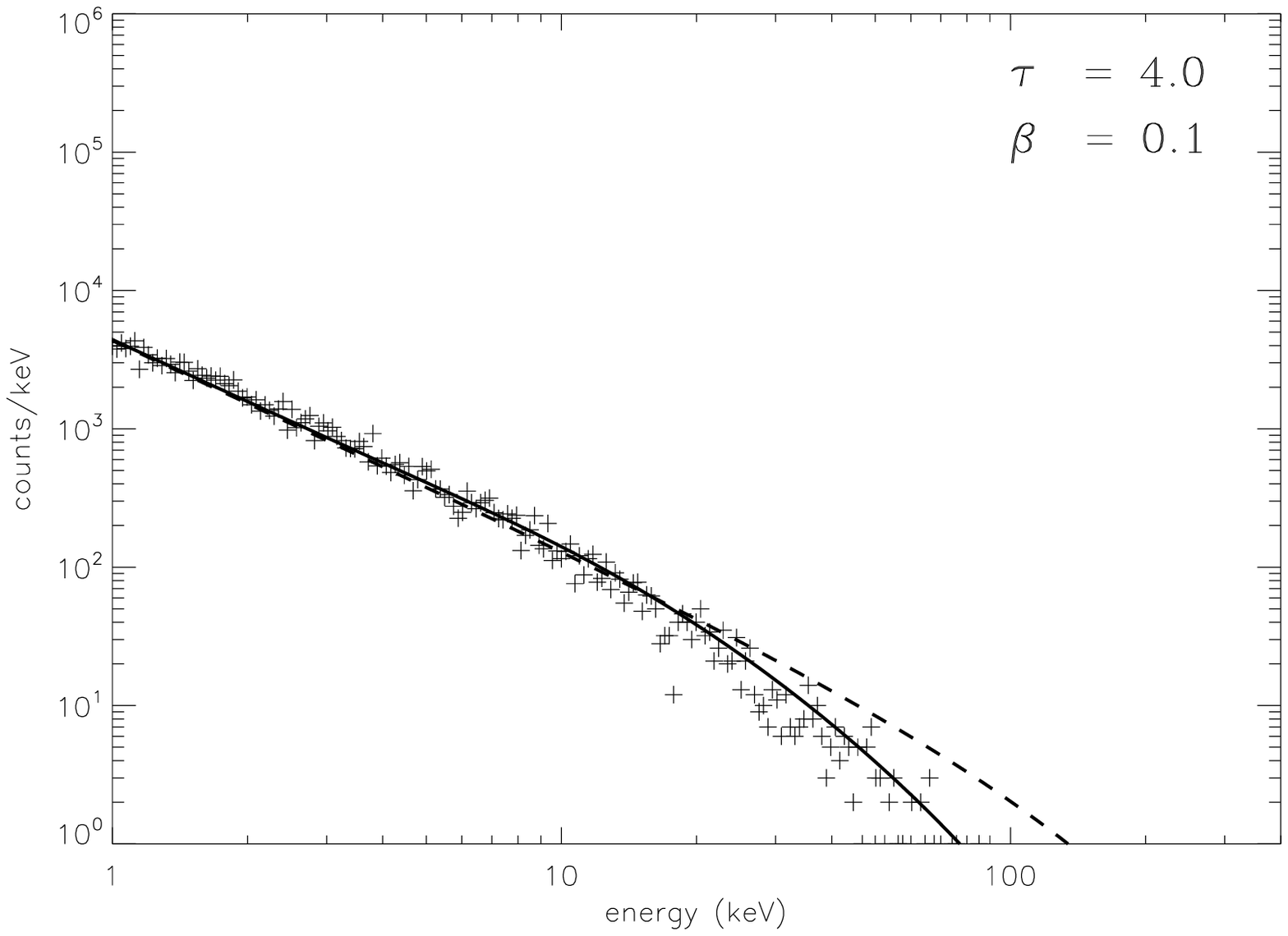}
\includegraphics[width=3.5in,height=3in]{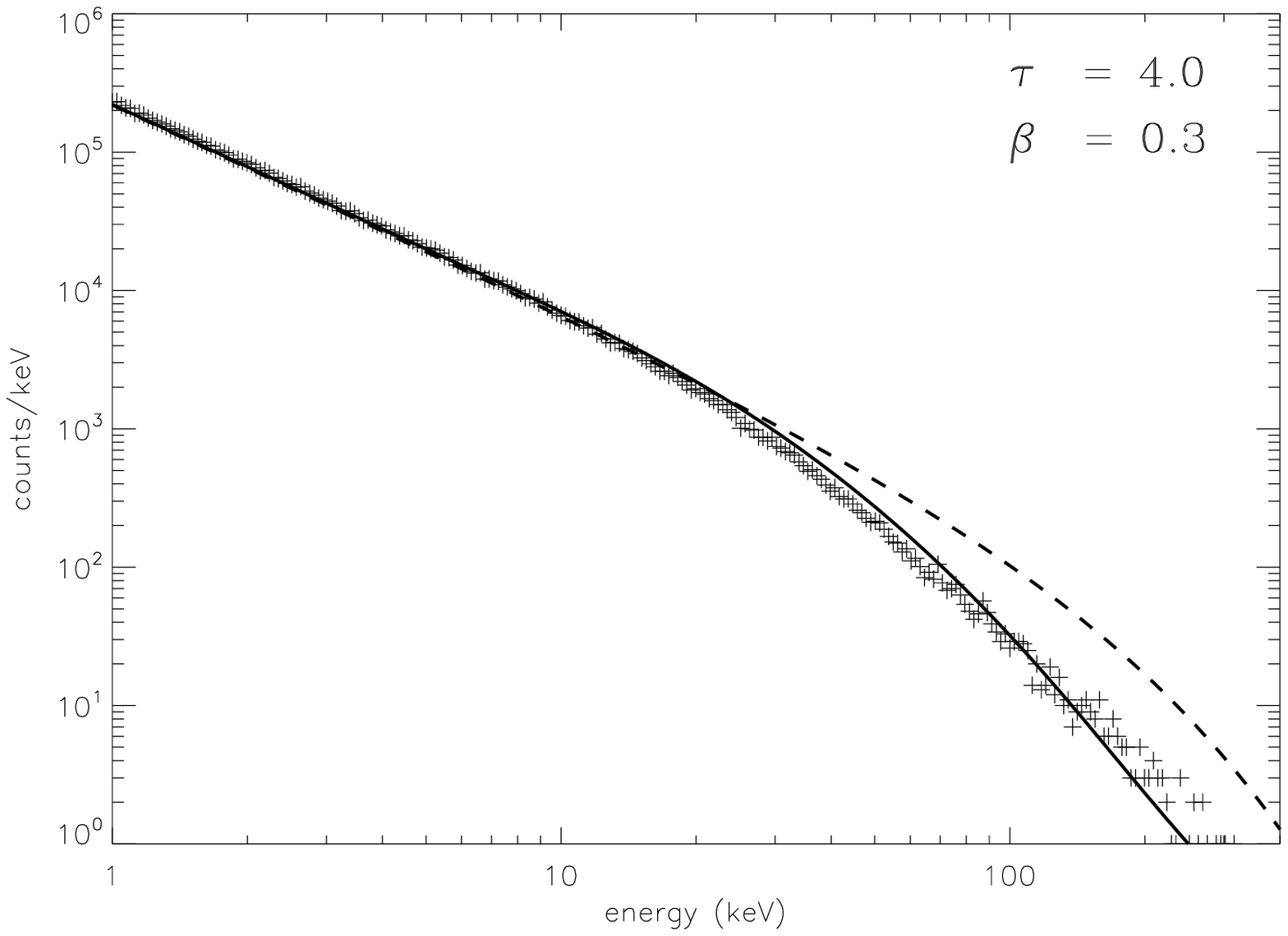}
\includegraphics[width=3.5in,height=3in]{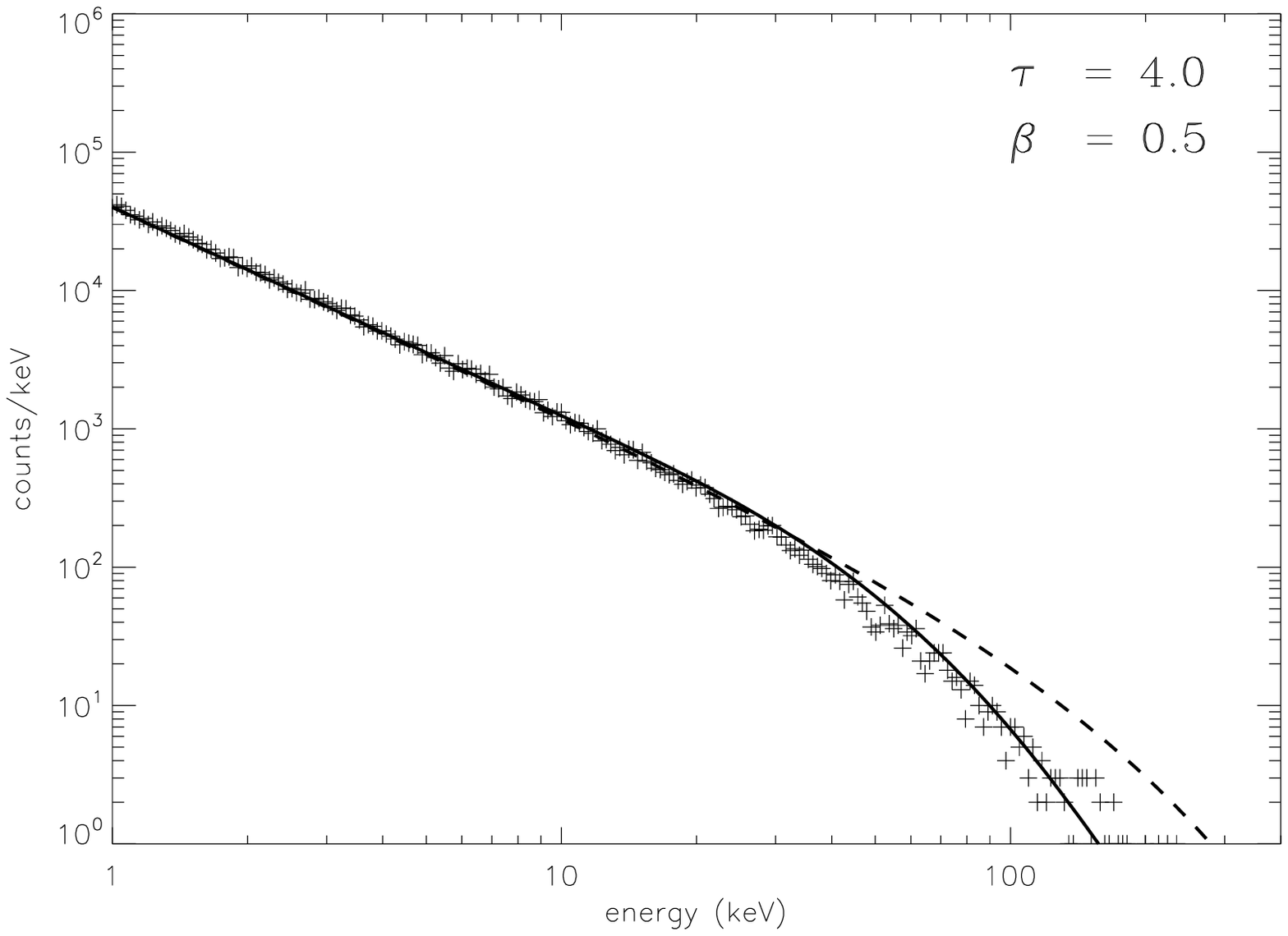}
\includegraphics[width=3.5in,height=3in]{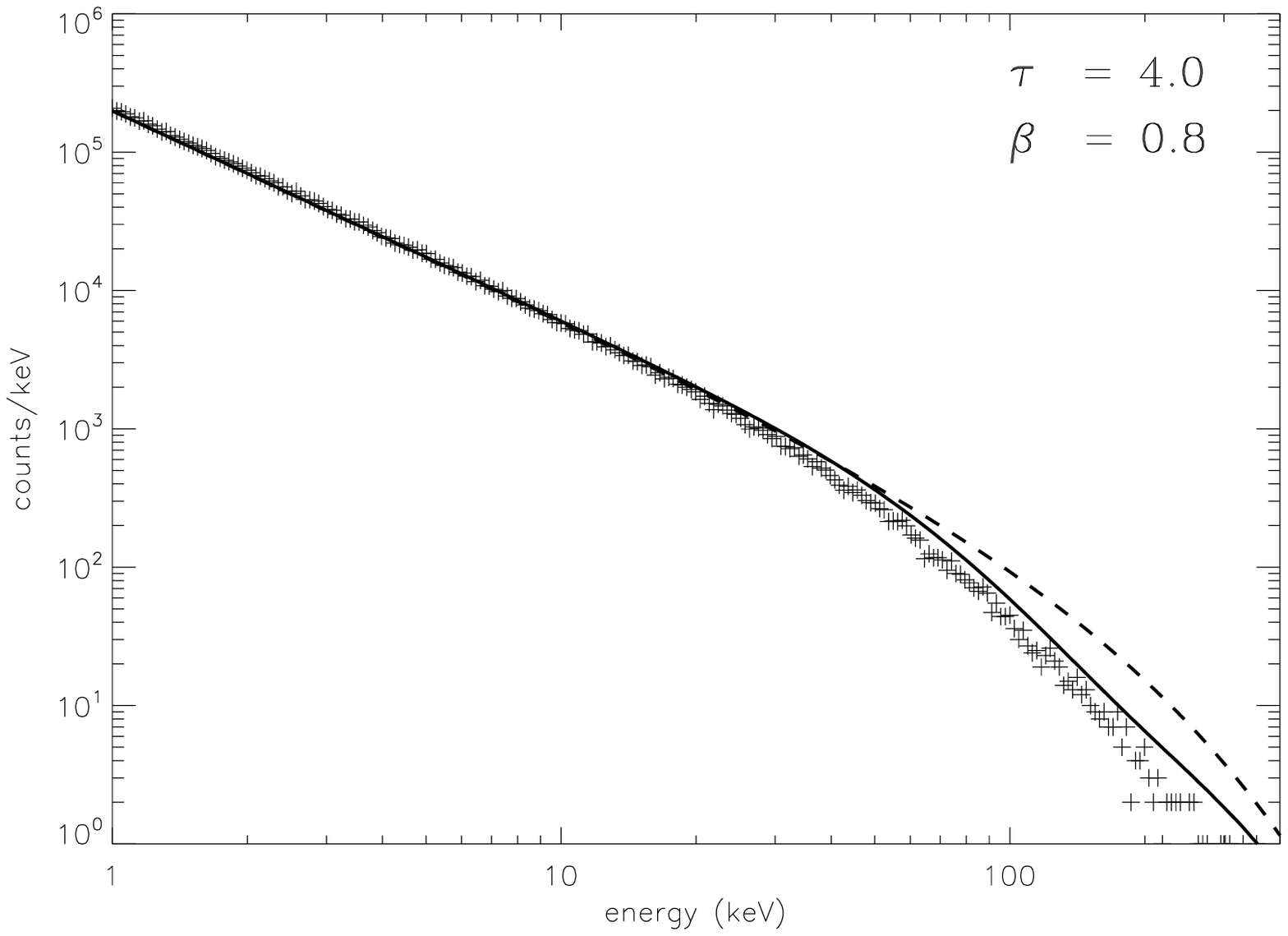}
\caption{Emerging spectrum of a compact object through an outflowing plasma. 
The incident spectrum inside the outflowing plasma is taken to be
$\varphi_{comp}(E) \propto E^{-\alpha}\exp(-E/E_{\ast})$ (see Eq. \ref{comp}) 
where energy index $\alpha=0.5$ and high energy cutoff at $E=E_{\ast}=130$ keV.
Outflow parameters are $kT_e=0.1$ keV, $\tau =4$.
 The photon emerging spectrum (cross points) is compared to the incident spectrum (dash line) and the analytical downscattering spectrum (solid line)
(see Eq. \ref{intexpan1}).  
For $\beta = 0.1,~0.3,~0.5,~0.8$, the best-fit parameters are $b=0.06,~0.10,~0.18,~0.38$ respectively.
{\it The low energy downscattering bump effect is not seen in the emergent spectra}.
}
\end{figure}

\newpage

\begin{figure}
\includegraphics[width=3.5in,height=3in]{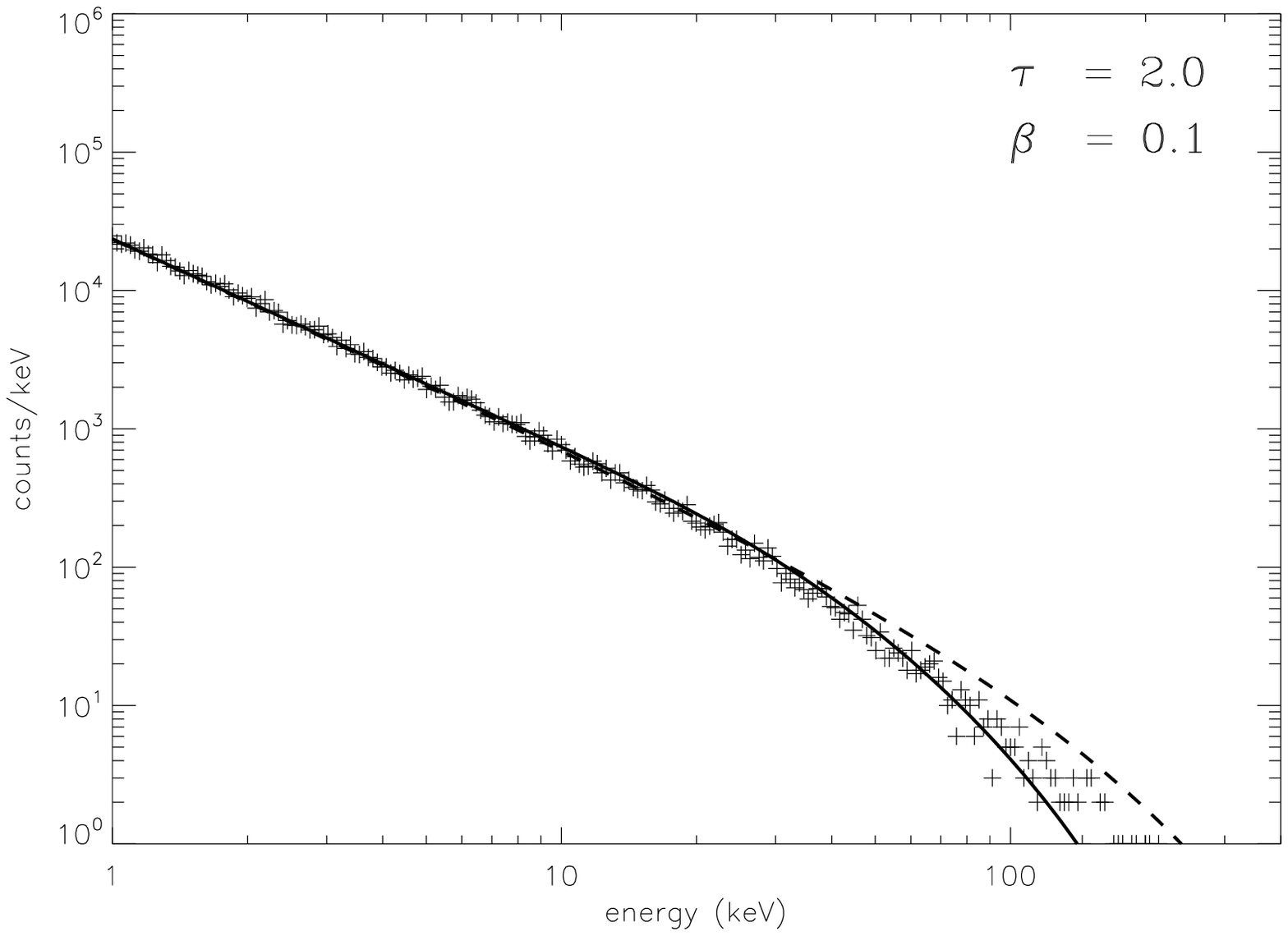}
\includegraphics[width=3.5in,height=3in]{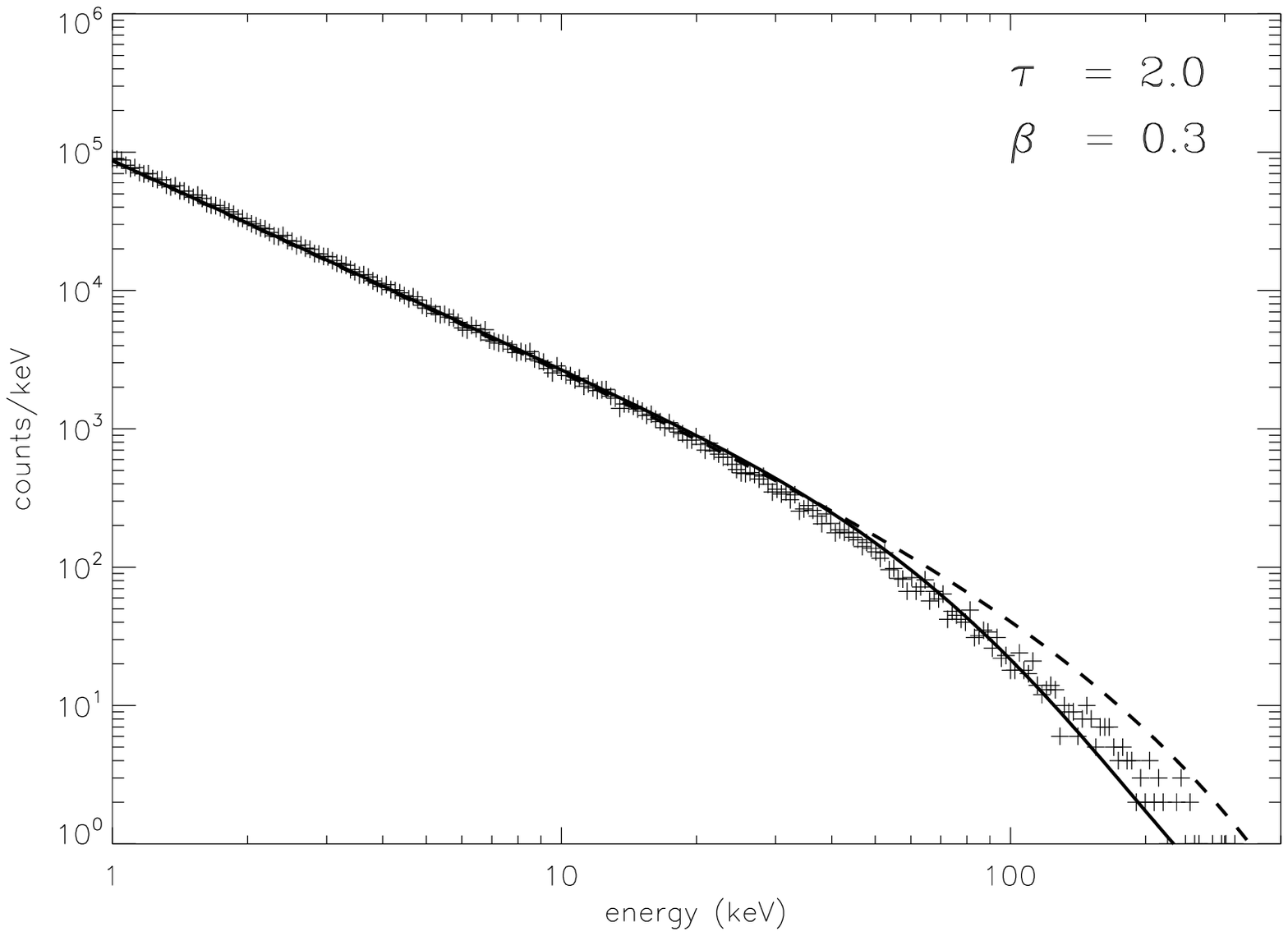}
\includegraphics[width=3.5in,height=3in]{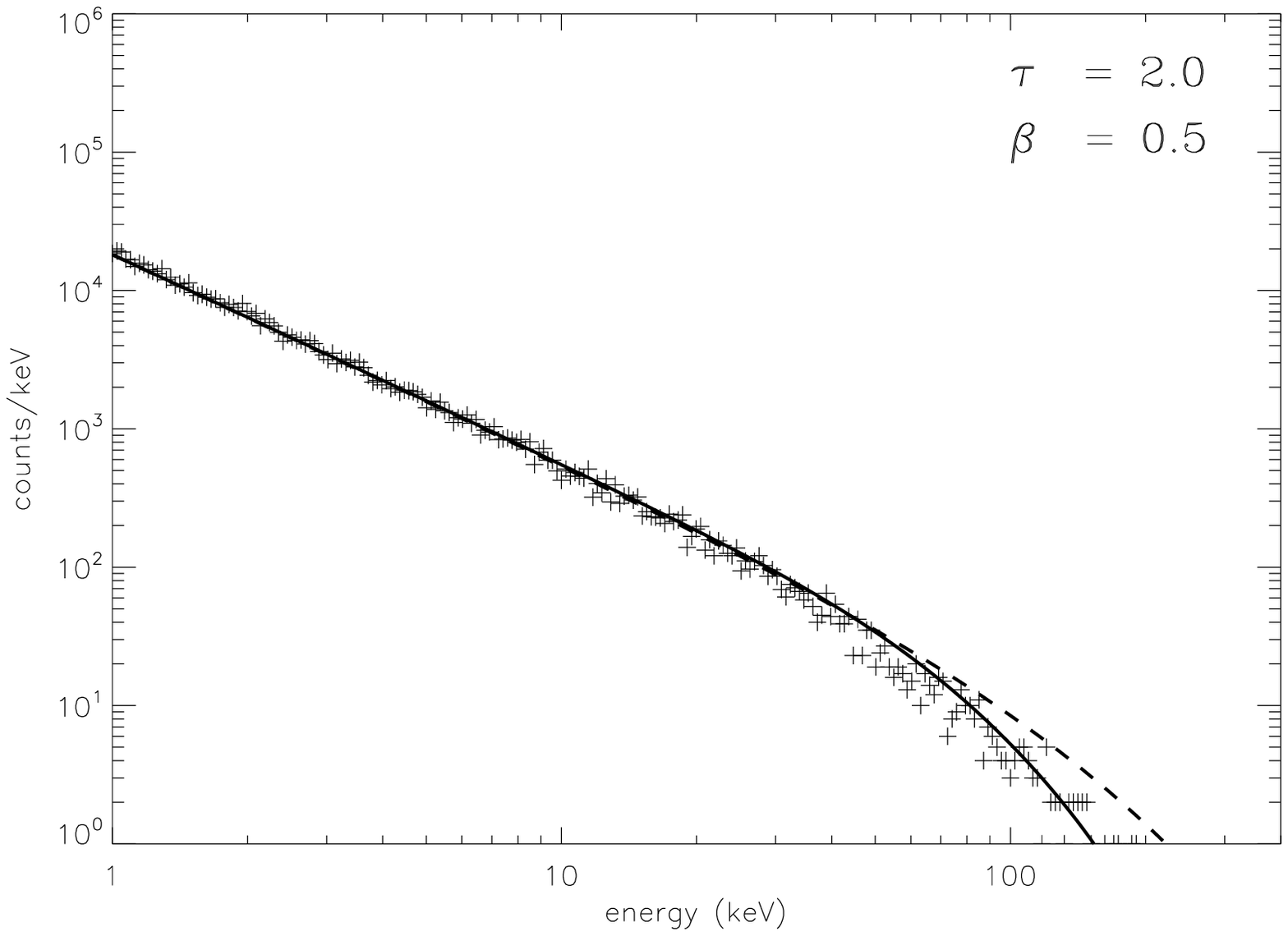}
\includegraphics[width=3.5in,height=3in]{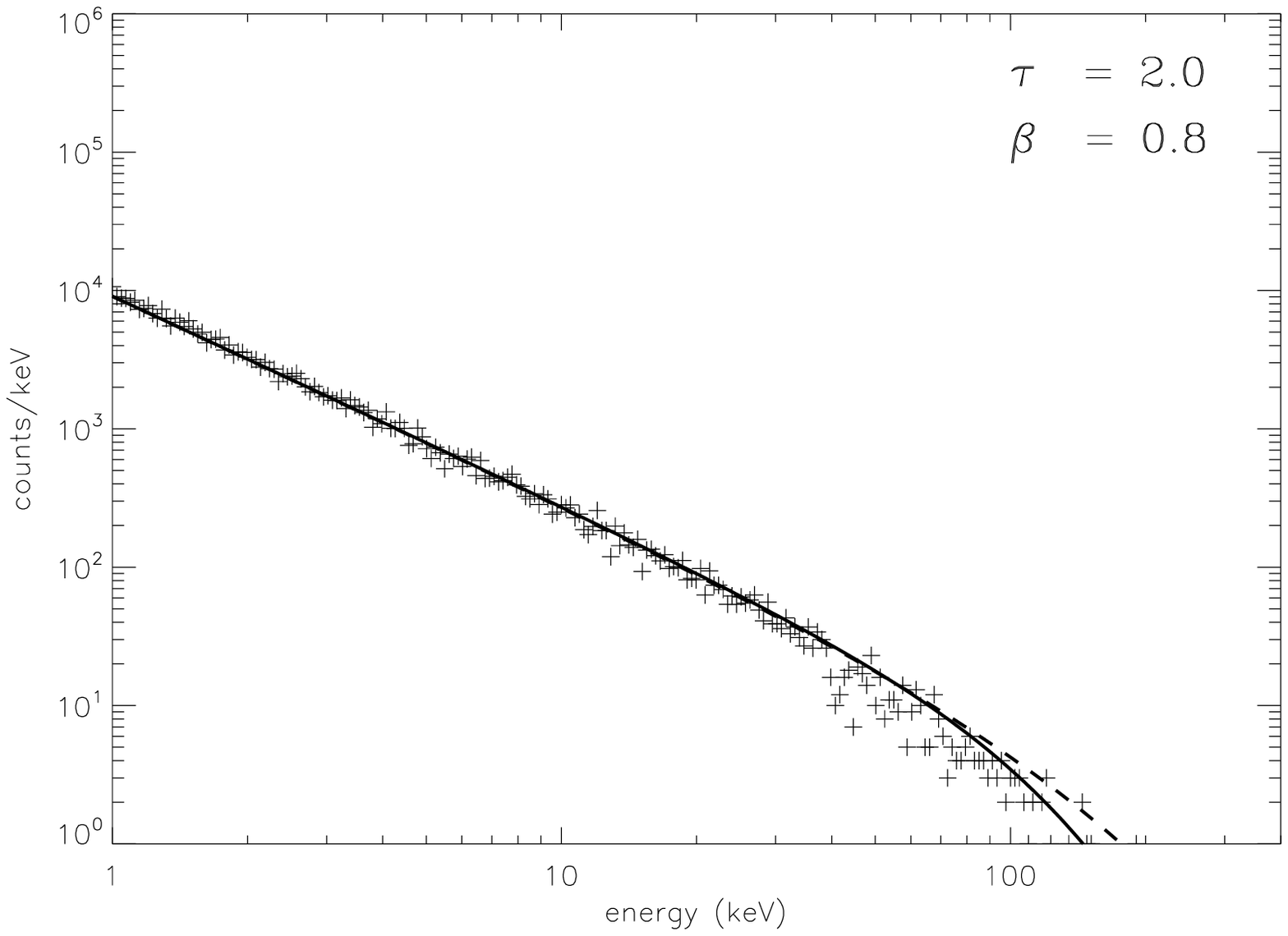}
\caption{The same as in Fig. 6 for $\tau=2$. For $\beta = 0.1,~0.3,~0.5,~0.8$, the best-fit parameters 
are $b=0.14,~0.21,~0.30,~0.59$ respectively. 
 }
\end{figure}

\newpage
\begin{figure}
\includegraphics[width=6.1in, height=6.8in,angle=90]{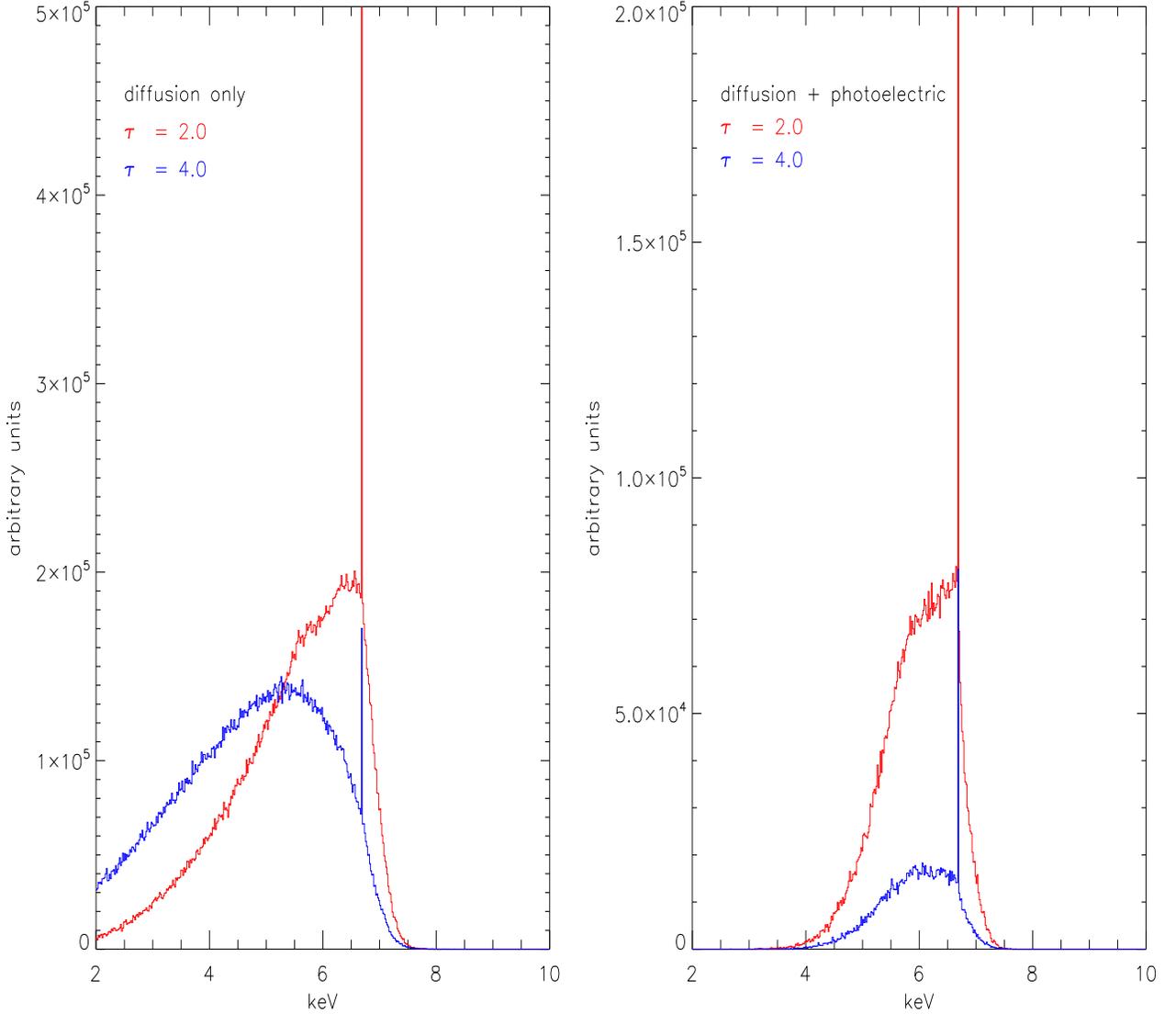}
\caption{Line profiles as a function of $\tau$ (for $\beta=0.1$). 
Primary photons are generated at the bottom of outflow. On the left-hand side : the line profiles for the pure scattering case
and on the right-hand side : that for scattering plus absorption case. In each case, the fixed parameters are  $E_{ph} = 6.6$ keV, $kT_e = 0.1$ keV. 
The normalization of the direct non-scattering line component is suppressed for $\tau=2$. The real normalizations  are 
 $A_N=1.35\times10^6, ~1.29\times10^6$ for the diffusion and diffusion+absorption cases respectively.
}
\end{figure}

\begin{figure}
\includegraphics[width=6.in,height=3.2in,angle=0]{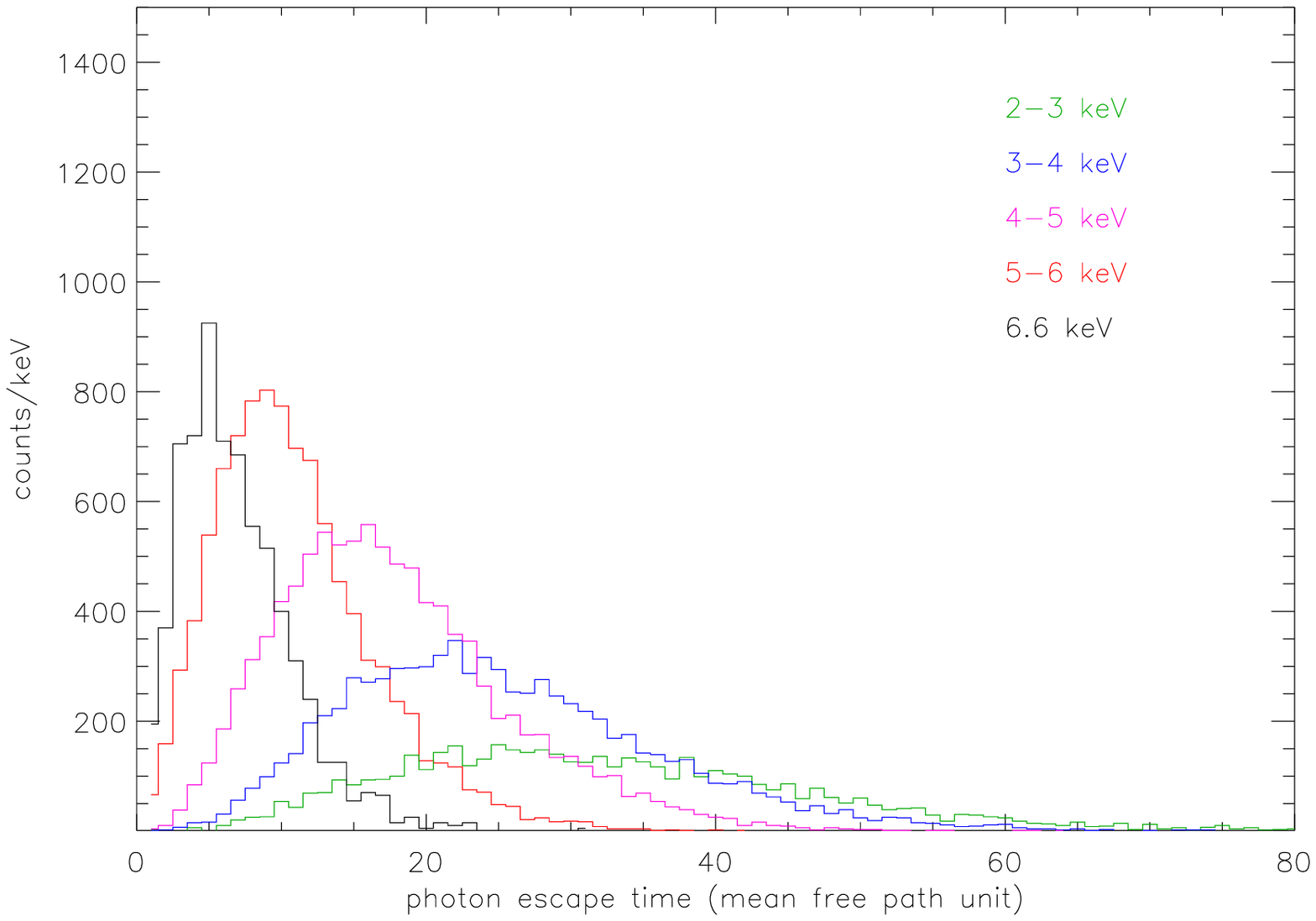}
\includegraphics[width=6.in,height=3.2in,angle=0]{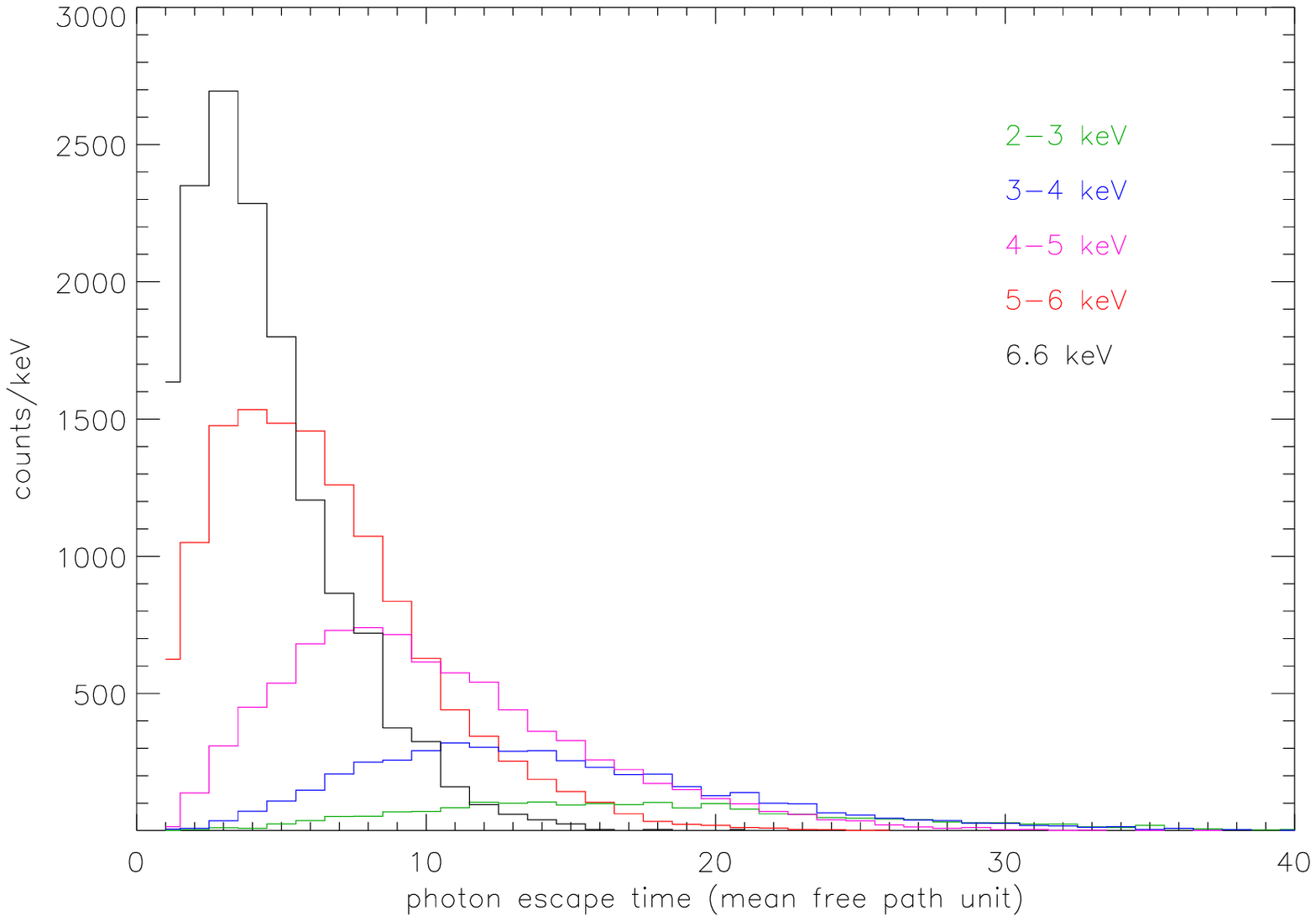}
\caption{Timing properties of the line propagation in the  outflow. 
A flash of monochromatic photons at 6.6 keV is simulated, and propagated. 
Photons escaping from the bottom atmosphere without scattering are at 6.6 keV (see Fig. 1). 
The model parameters used here are  
$kT_e = 0.1$ keV, ~$ \beta = 0.1$ and 
$\tau =4$ (upper panel),   and  $\tau=2$ (lower panel).
We show the arrival time of photons at the top of the cloud for five energy bands, 
in free path  time units $t_{fp}=l/c$.  Soft lags, low energy emission at later times are clearly seen. 
The photon distribution  over escape time integrated over energies can be fitted  with an exponential 
law  $b\exp(-bt/t_{fp})$ (shown in Fig.  2). The values of $b=0.057$ and $b=0.144$ for $\tau=4$ and $\tau=2$ respectively 
(and consequently average number of scatterings $N_{av}=1/b$) are very close to $b$ (or $N_{av}$) obtained 
from the best-fit analytical spectrum for the same values of $\beta$ and $\tau$  (see text).} 
\end{figure}

\begin{figure}
\includegraphics[width=3.1in,height=6.8in,angle=90]{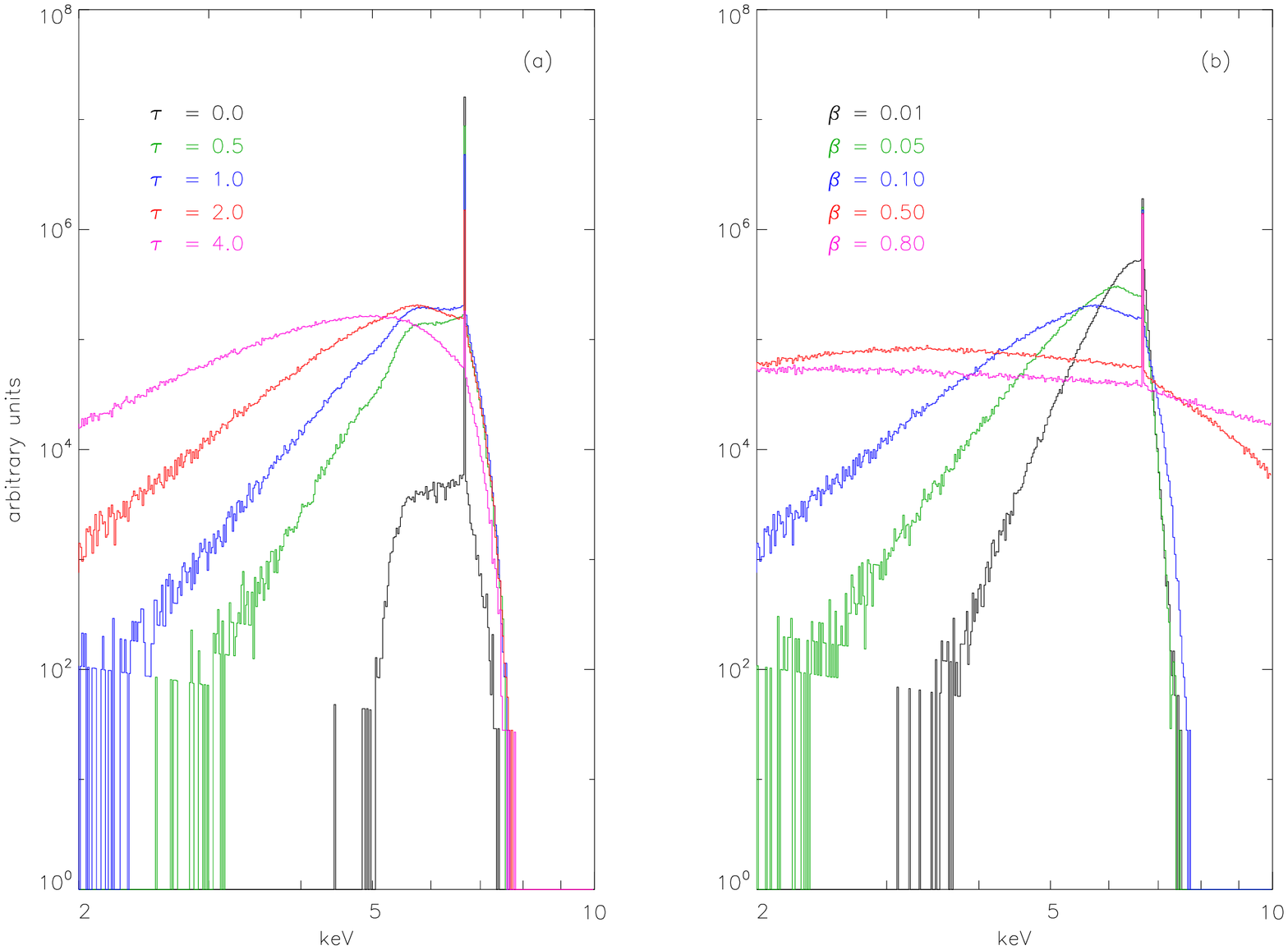}
\includegraphics[width=3.1in,height=6.8in,angle=90]{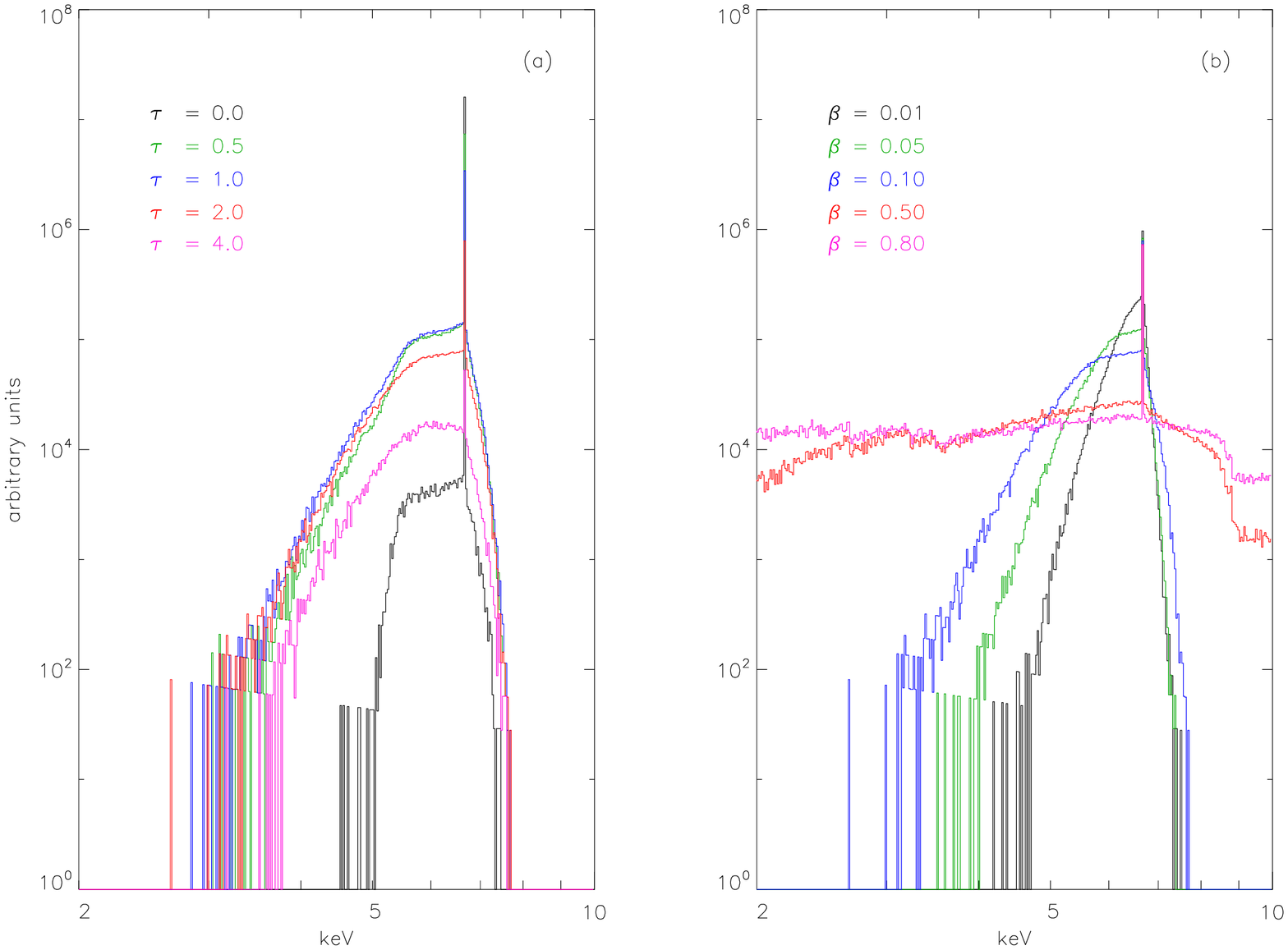}
\caption{Line profiles as a function of $\tau$ and $\beta$. 
Primary photons are generated in the bottom of outflow. 
Upper panel: the line profiles for the pure scattering case. 
Lower panel:  that for scattering plus absorption case. 
In each case, the fixed parameters are  $E_{ph} = 6.6$ keV and  $kT_e = 0.1$ keV. Spectral lines as a function of $\tau$ are for $\beta=0.1$ and that as a function of
$\beta$ for $\tau=2$. Note, the case with a notation $\tau=0.0$ [see upper and lower panels (a)], corresponds to $\tau=0.01$. 
}
\end{figure}
\begin{figure}
\includegraphics[width=7.in,height=4.5in,angle=0]{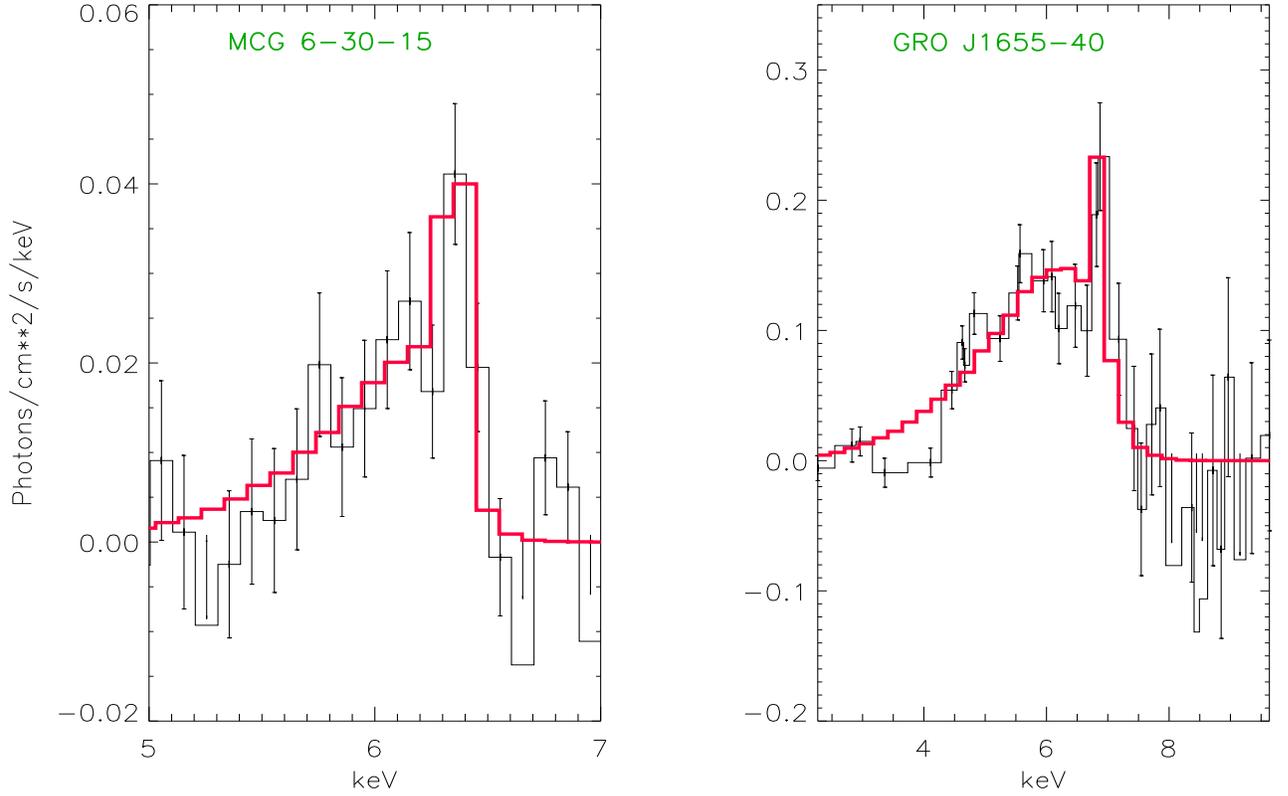}
\caption{Fits of our model to  the XMM observations of MCG 6-30-15 
(Wilms et al. 2003) and  to the ASCA observations of GRO J1655-40 (Miller et al. 2004b).
For MCG 6-30-15, we have fitted the continuum with a power law of 
index 1.8 and compare the residuals with pure scattering model. 
The pure scattering model. 
For this model the best-fit parameters which we found are $\rm E_{ph} = 6.51$ keV, ~$kT_e = 0.1$ keV, 
$\tau =1.2, ~\beta = 0.02$. Primary photons are generated at the bottom of outflow (illumination of outflow
from inside).
For GRO J1655-40 we found the following  best-fit parameters: $\rm E_{ph} = 7.1$ keV, ~$kT_e = 0.1$ keV, 
$\tau =2.0, ~\beta = 0.1$.}
  
\end{figure}

\begin{figure}
\includegraphics[width=7.in,height=4.5in,angle=0]{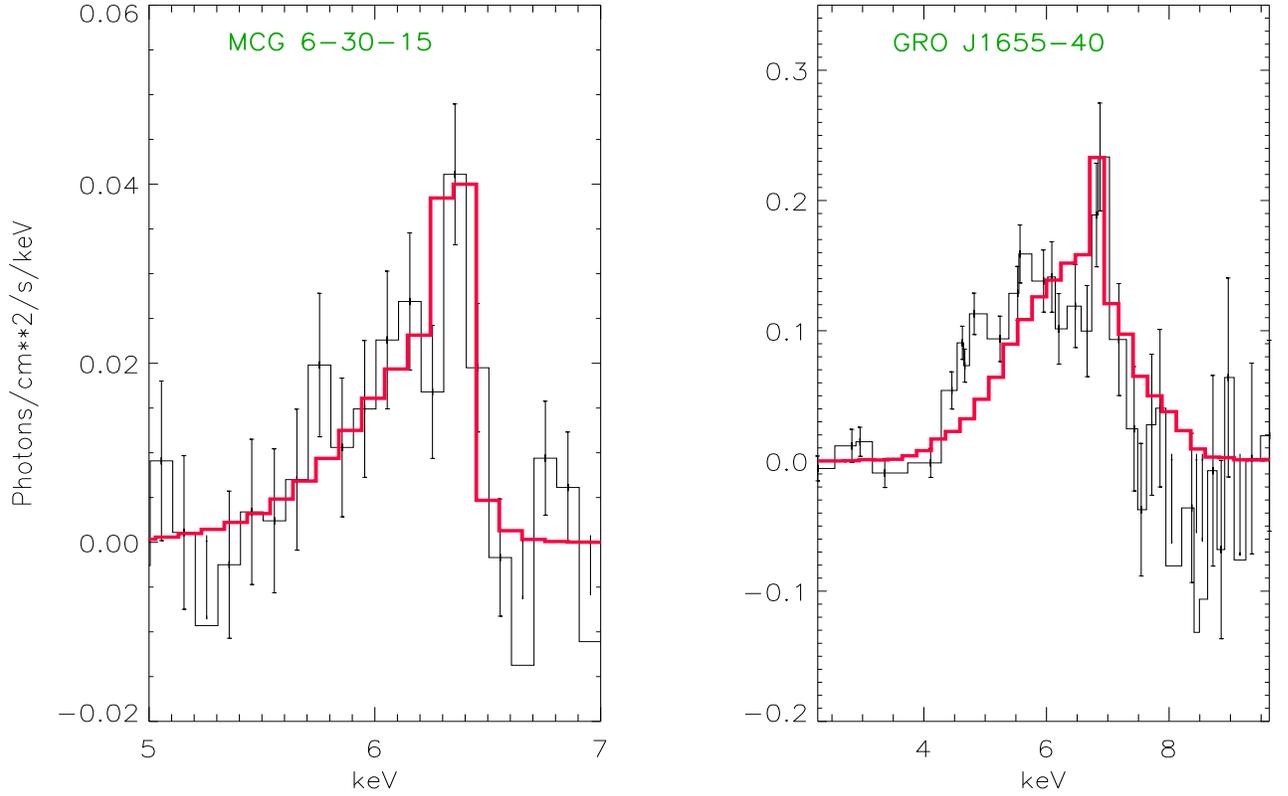}
\caption{Fits of our model to  the XMM observations of MCG 6-30-15 
 and  to the ASCA observations of GRO J1655-40 using the model for which the photo-absorption along  with scattering
 are included.
For MCG 6-30-15 the best-fit model parameters which we found are $\rm E_{ph} = 6.51$ keV, ~$kT_e = 0.1$ keV, 
$\tau =1.7, ~\beta = 0.02$. 
For GRO J1655-40 
 we found the following best-fit parameters: $\rm E_{ph} = 7.1$ keV, ~$kT_e = 0.1$ keV, 
$\tau =3.3, ~\beta = 0.25$.} 
\end{figure}

\begin{figure}
\includegraphics[width=3.1in,height=3.2in,angle=0]{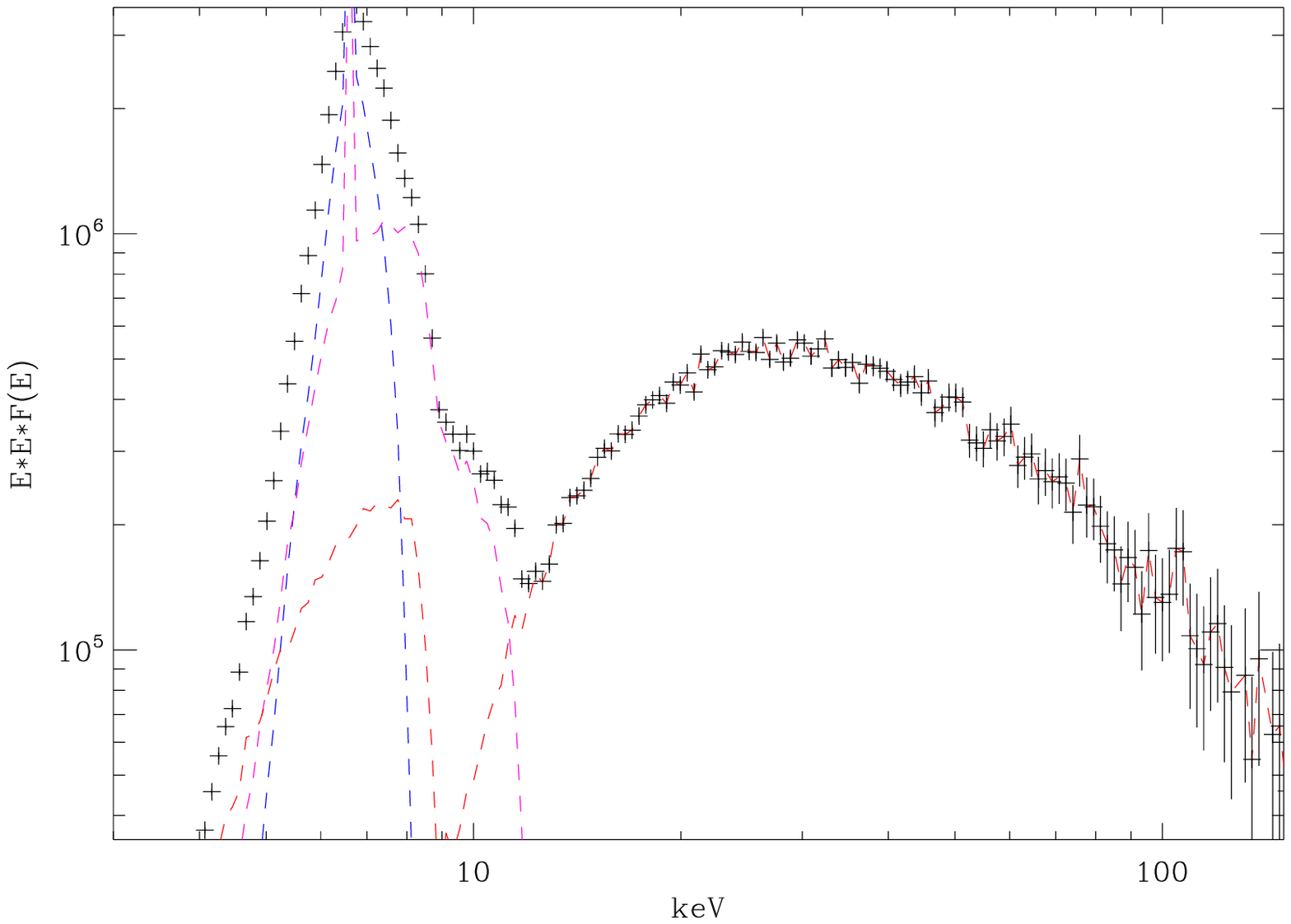}
\includegraphics[width=3.1in,height=3.2in,angle=0]{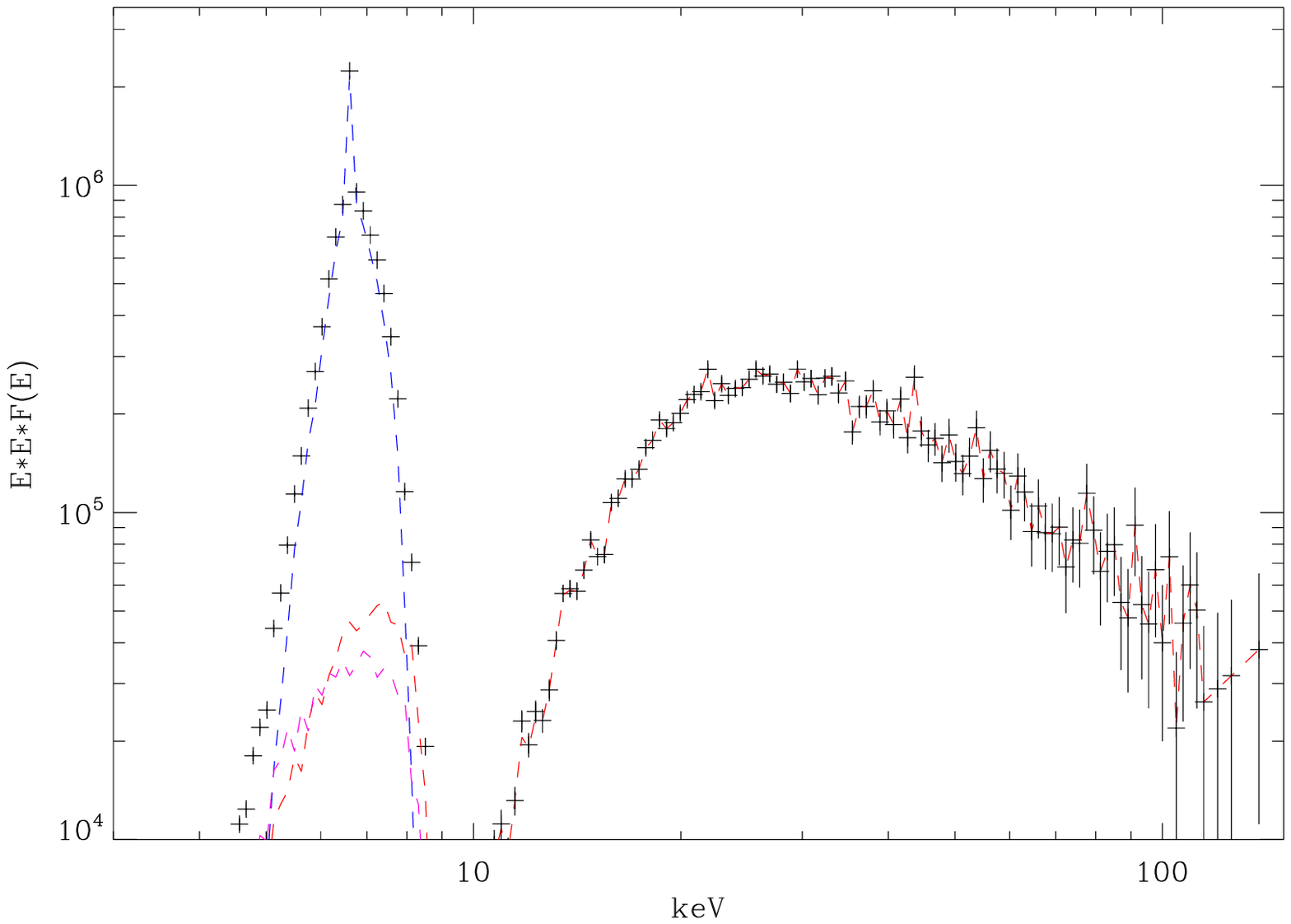}
\includegraphics[width=3.1in,height=3.2in,angle=0]{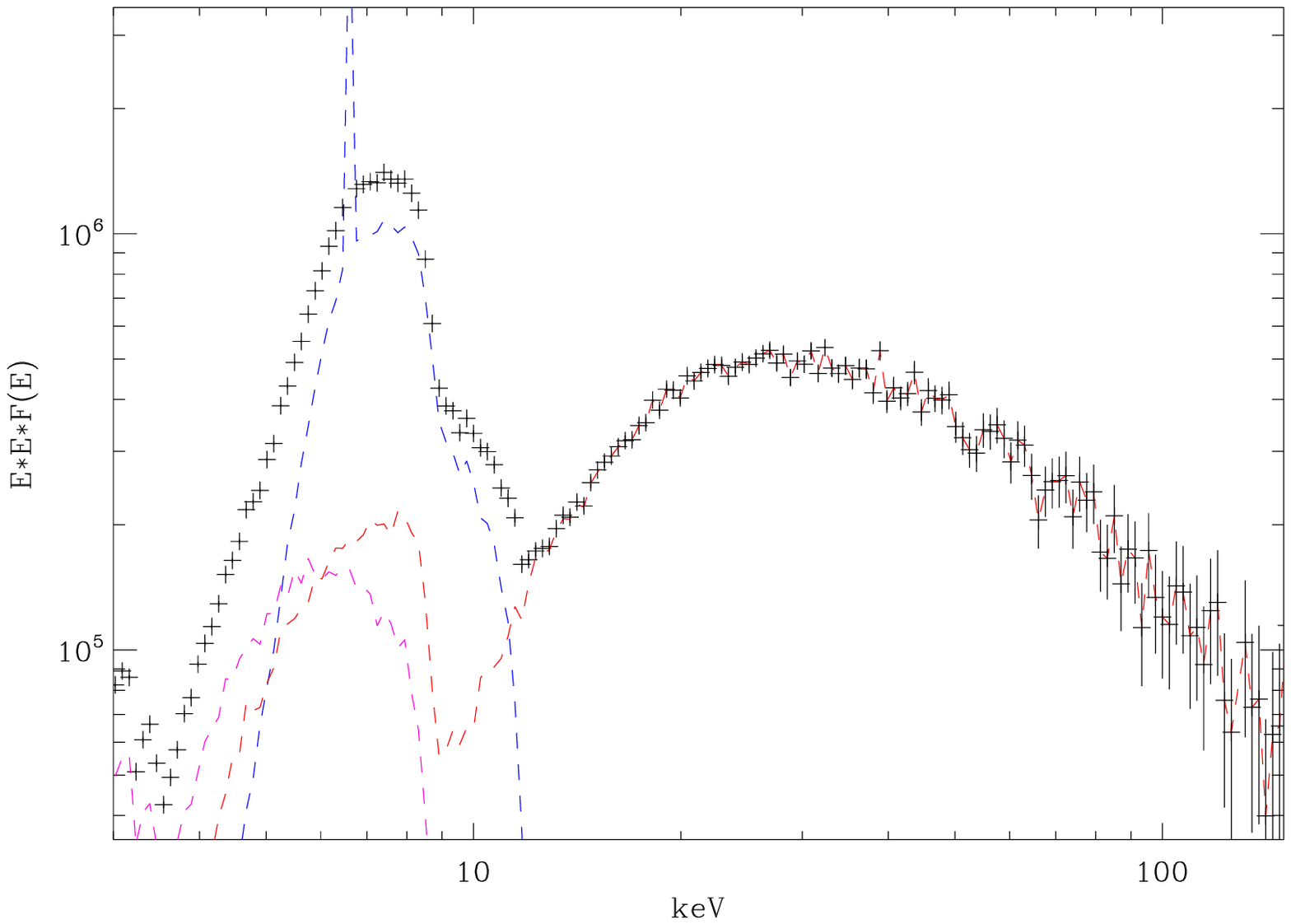}
\includegraphics[width=3.1in,height=3.2in,angle=0]{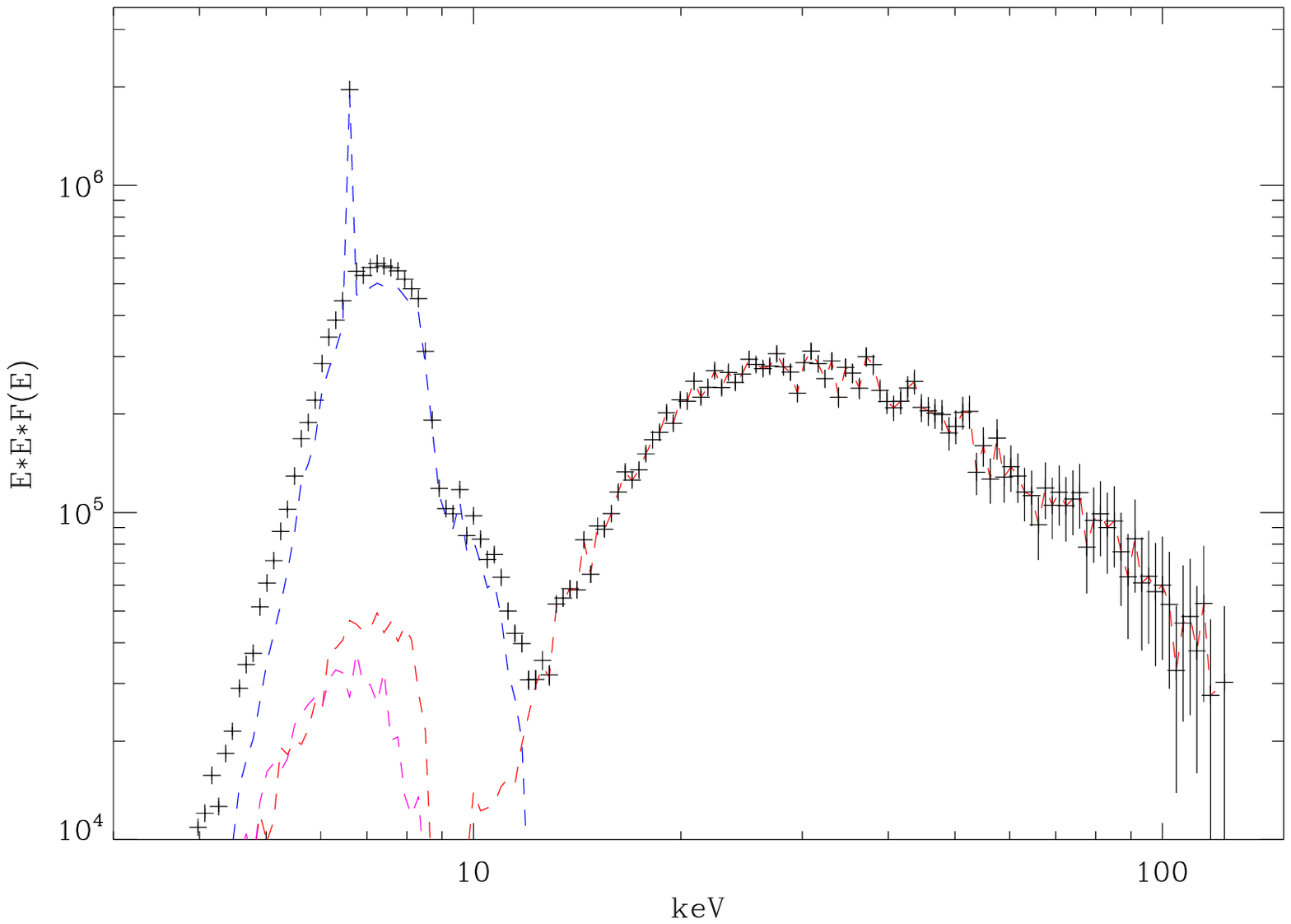}
\caption{The central source spectrum reprocessed through the wind shell.
Upper panel: $\beta=0.1$ and $\tau_0=2$ and $\tau_0=4$. Lower panel: that is for $\beta=0.3$. 
In the X-ray spectrum of the central source the photon numbers are the same in the blackbody and hard
components. In fact, the ratio of the photon numbers is arbitrary and 
it depends on the illumination of the Compton cloud by the source of blackbody radiation.
In this simulations we assume the blackbody color temperature is 1.2 keV.
The resulting spectrum is shown by black histogram, whereas orange and pink  curves present the  hard and blackbody components of the central source 
reprocessed in the wind respectively. A blue curve is the K$_{\alpha}$ line formed in the wind. One can clearly see the strong fluorescent K$_{\alpha}$ line, K-edge formed in the wind and a prominent bump
around 25 keV. The line  consists of narrow and broad components (see narrow blue peak at 6.4 keV in all spectra). 
}
\end{figure}

\end{document}